\documentclass{article}
\usepackage{graphics,latexsym,amssymb,amsmath}
\usepackage{natbib}
\RequirePackage{graphics,epsfig,psfrag}
\usepackage{amsthm}
\usepackage{hyperref}
\hypersetup{
    bookmarks=true,         % show bookmarks bar?
    unicode=false,          % non-Latin characters in Acrobats bookmarks
    pdftoolbar=true,        % show Acrobats toolbar?
    pdfmenubar=false,        % show Acrobats menu?
    pdffitwindow=false,     % window fit to page when opened
    pdfstartview={FitH},    % fits the width of the page to the window
    pdftitle={Computing Chebyshev knot diagrams},    % title
    pdfauthor={PVK, DP, FR, CT},     % author
    pdfsubject={},   % subject of the document
    pdfcreator={PVK},   % creator of the document
    pdfproducer={Producer}, % producer of the document
    pdfkeywords={keyword1} {key2} {key3}, % list of keywords
    pdfnewwindow=true,      % links in new window
    colorlinks=true,       % false: boxed links; true: colored links
    linkcolor=blue,          % color of internal links
    citecolor=blue,        % color of links to bibliography
    filecolor=magenta,      % color of file links
    urlcolor=cyan           % color of external links
}
\newcommand\Prod{\mathop{\prod}\limits}
\def\abs#1{\vert #1 \vert}
\def\Abs#1{\left \vert #1 \right \vert}

\def\norm#1{\vert \vert #1 \vert \vert}
\def\Mod#1{\,(\hbox{\rm mod}\,#1)}
\def\RR{{\bf R}} %reelle Zahlen
\def\NN{{\bf N}} %reelle Zahlen
\def\QQ{{\bf Q}} %reelle Zahlen
\def\ZZ{{\bf Z}} %reelle Zahlen
\def\CC{{\bf C}}
\def\cE{{\cal E}}

\def\undemi{\hbox{\raise2pt\hbox{\tiny $1$}\hskip-0.5pt/\hskip-0.5pt\raise-1pt\hbox{\tiny $2$}}}
\def\SS{{\bf S}} %reelle Zahlen
\newcommand\Disc{\mathrm{Disc}\,}
\newcommand\bn{\begin{enumerate}}
\newcommand\en{\end{enumerate}}
\def\pn{\medskip\par\noindent}
\def\Frac#1#2{{\displaystyle{{#1}\over{#2}}}}

\def\[#1\]{\begin{equation}#1\end{equation}}
\def\$#1\${\begin{eqnarray}#1\end{eqnarray}}

\def\phi{\varphi}

\def\mod{\hbox{\rm mod}\,}
\def\Res{\hbox{\rm Res}}
\newcommand\cO{\mathcal{O}}

\newcommand\tcO{\widetilde\cO}
\def\bi{\vspace{-4pt}\begin{itemize}\itemsep -3pt plus 1pt minus 1pt}
\def\ei{\end{itemize}\vspace{-4pt}}
\def\cC{{\cal C}}

\def\cZ{{\cal Z}}
\def\cV{{\cal V}}

\def\pent#1#2{\pe{\frac{#1}{#2}}}
\def\pent#1#2{\lfloor{\frac{#1}{#2}}\rfloor}

\def\sign#1{{\mathrm{sign}}\,#1}
\def\taut{\tau_{\cT}}
\newcommand{\Pf}{\pn{\em Proof}. }

\newcommand{\EPf}{\hbox{}\hfill$\Box$\vspace{.5cm}}

\def\abs#1{\left \vert #1 \right \vert}
\def\Frac#1#2{{\displaystyle{{#1} \overwithdelims.. {#2}}}}
\def\frac#1#2{{\textstyle{{#1} \overwithdelims.. {#2}}}}
\def\phi{\varphi}

\newtheorem{thm}{Theorem}[section]
{\theoremstyle{definition}\newtheorem{defn}[thm]{Definition}}
{\theoremstyle{definition}\newtheorem{rem}[thm]{Remark}}
\newtheorem{lem}[thm]{Lemma}
\newtheorem{prop}[thm]{Proposition}
\newtheorem{cor}[thm]{Corollary}
\newtheorem{thm*}[thm]{Theorem}

\date{\today}

\begin{document}
\pagestyle{myheadings}
\markboth{P. -V. Koseleff, D. Pecker, F. Rouillier \quad \today}{{\em Computing Chebyshev knot diagrams, draft version, \today}}
%%%%%%%%%%%%%%%%%%%%% Publisher's Area please ignore %%%%%%%%%%%%%%
%wis%\catchline{}{}{}{}{}
%%%%%%%%%%%%%%%%%%%%%%%%%%%%%%%%%%%%%%%%%%%%%%%%%%%%%%%%%%%%%%%%%%%
\title{Computing Chebyshev knot diagrams}% \\
\author{P. -V. Koseleff,  D. Pecker, F. Rouillier, C. Tran}
\maketitle
\begin{abstract}
A Chebyshev curve  $\cC(a,b,c,\phi)$ has a parametrization of the form
$ x(t)=T_a(t)$; \  $y(t)=T_b(t)$; $z(t)= T_c(t + \phi)$,  where $a,b,c$
are integers, $T_n(t)$ is the Chebyshev polynomial
%(henceforward we briefly call Chebyshev polynomial)
of degree $n$ and $\phi \in \RR$. When $\cC(a,b,c,\phi)$ is nonsingular,
it defines a polynomial knot. 
We determine all possible knot diagrams when $\phi$ varies. 
Let $a,b,c$ be integers, $a$ is odd, $(a,b)=1$, we show that 
one can list all possible knots $\cC(a,b,c,\phi)$ in
$\tcO(n^2)$ bit operations, with $n=abc$. 
\end{abstract}
\begin{small}
\tableofcontents
\end{small}
%%%%%%%%%%%%%%%%%%%%%%%%%%
%\modulolinenumbers[5]
%\linenumbers
%%%%%%%%%%%%%%%%%%%%%%%%%%%%%%%%%%%%%%%%%%%%%%%%%%%%%%%%%%%%%%%%%%%%%%%%%%%%%%%
\section{Introduction}\label{intro}

It is known that every knot in $\SS^3$ can be represented as the closure of the
image of a polynomial embedding $\RR \to \RR ^3 \subset \SS^3 $, see \citet{Va}.
Given a knot $K$, it is in general a difficult problem
to determine $(a,b,c)$ such that there exists a polynomial embedding
$\RR \to \RR^3$ of multi-degree $(a,b,c)$ that parametrizes $K$ and an even more difficult problem to determine a {\em minimal} $(a,b,c)$,
for the lexicographic ordering,
see for example \citep*{BKP}.
\pn
A Chebyshev curve $\cC(a,b,c,\varphi)$ is the space curve
$$ \cC(a,b,c,\varphi) : x=T_a(t), \, y=T_b(t), \, z=T_c(t+\varphi),$$
where $T_n(x) = 2 \cos(n \arccos x/2)$ is the monic Chebyshev polynomial
of degree $n$, $a$, $b$, $c$ with $a,b$
coprime and $a<b$, are positive integers, and $\varphi$ is a real number.
If a Chebyshev curve  $\cC(a,b,c,\varphi)$ is
nonsingular, then it defines a (long) knot.
\pn
Chebyshev knots are polynomial analogues of Lissajous knots,
which admit parametrizations of the form
$x=\cos (at); y=\cos(bt+\varphi); z=\cos(ct+ \psi)$.
These knots were first introduced by \citet*{BHJS}. It is known that every
knot is not necessarily a Lissajous knot.
Recently, it is shown in \citep{SV16} that every knot is a Fourier knot,
which admits a parametrization of the form
$x=\cos (at); y=\cos(bt+\varphi); z= \lambda \cos(ct+ \psi) +
\lambda' \cos (c't+\psi')$.
\pn
In \citep{KP3}, it is proved that every (long) knot
$K \subset \RR^3 \subset \SS^3$ is a Chebyshev knot,
that is to say there exists a Chebyshev curve $\cC(a,b,c,\varphi)$
that is isotopic to $K$ in $\SS^3$.
\pn
The objective of our contribution is to compute minimal Chebyshev
parametrization exhaustively for the first two-bridge knots
with 10 crossings and fewer.

Our strategy consists in studying exhaustively Chebyshev curves with
increasing degrees and identify the knots they represent. As every
knot is a Chebyshev knot, this process will describe all the
first knots as soon as we can identify each knot. To identify a knot,
we compute its diagram and this computation is the core of the process.

\subsection{Chebyshev Diagrams}

\begin{figure}[!ht]
\begin{center}
\begin{tabular}{ccc}
\epsfig{file=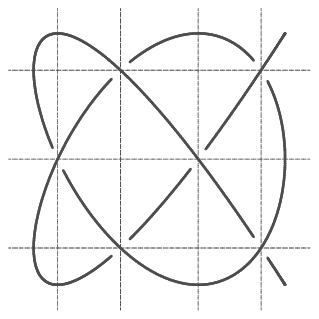,height=2.5cm,width=2.5cm}&
\epsfig{file=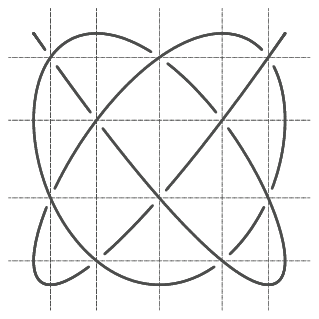,height=2.5cm,width=2.5cm}&
\epsfig{file=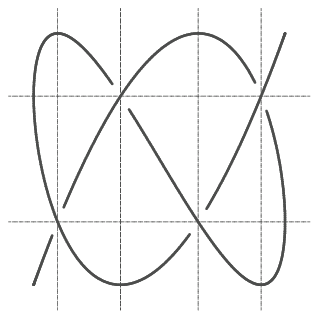,height=2.5cm,width=2.5cm}\\
$\overline{5}_2$&${5}_2$&$\overline{4}_1$\\
\epsfig{file=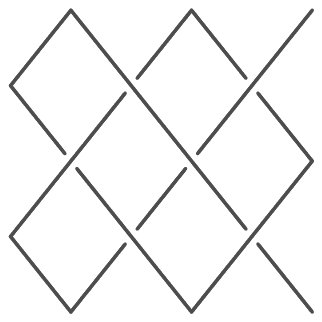,height=2.5cm,width=2.5cm}&
\epsfig{file=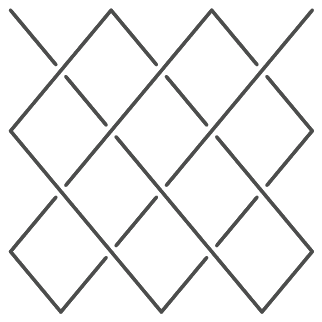,height=2.5cm,width=2.5cm}&
\epsfig{file=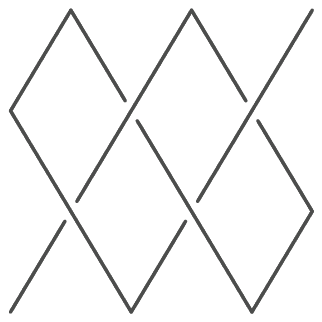,height=2.5cm,width=2.5cm}
\end{tabular}
\caption{Some Chebyshev knot diagrams and their billiard trajectories}\label{bt}
\end{center}
\end{figure}

\begin{figure}[!ht]
\begin{center}
\begin{tabular}{ccc}
{\scalebox{.1}{\includegraphics{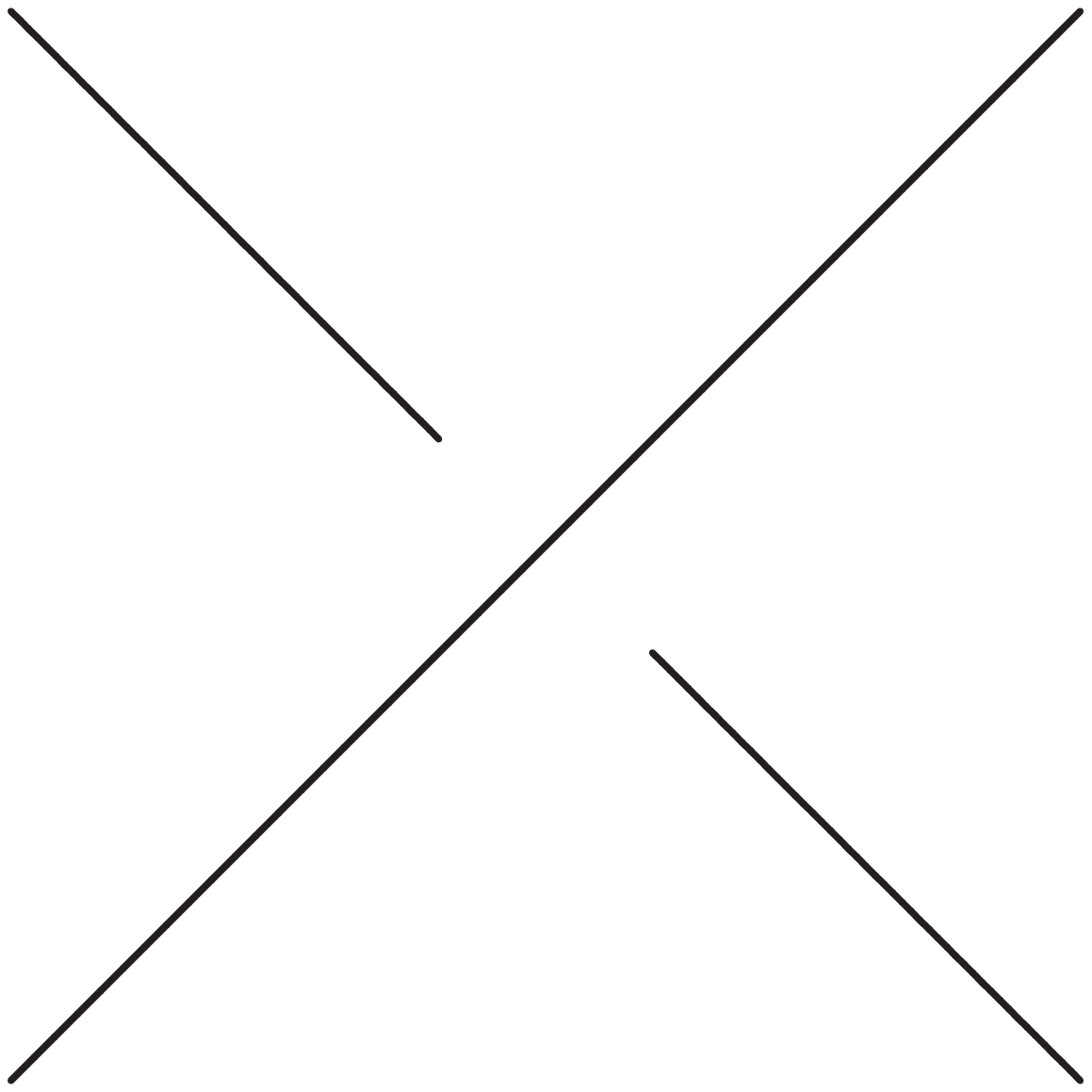}}} &\quad&
{\scalebox{.1}{\includegraphics{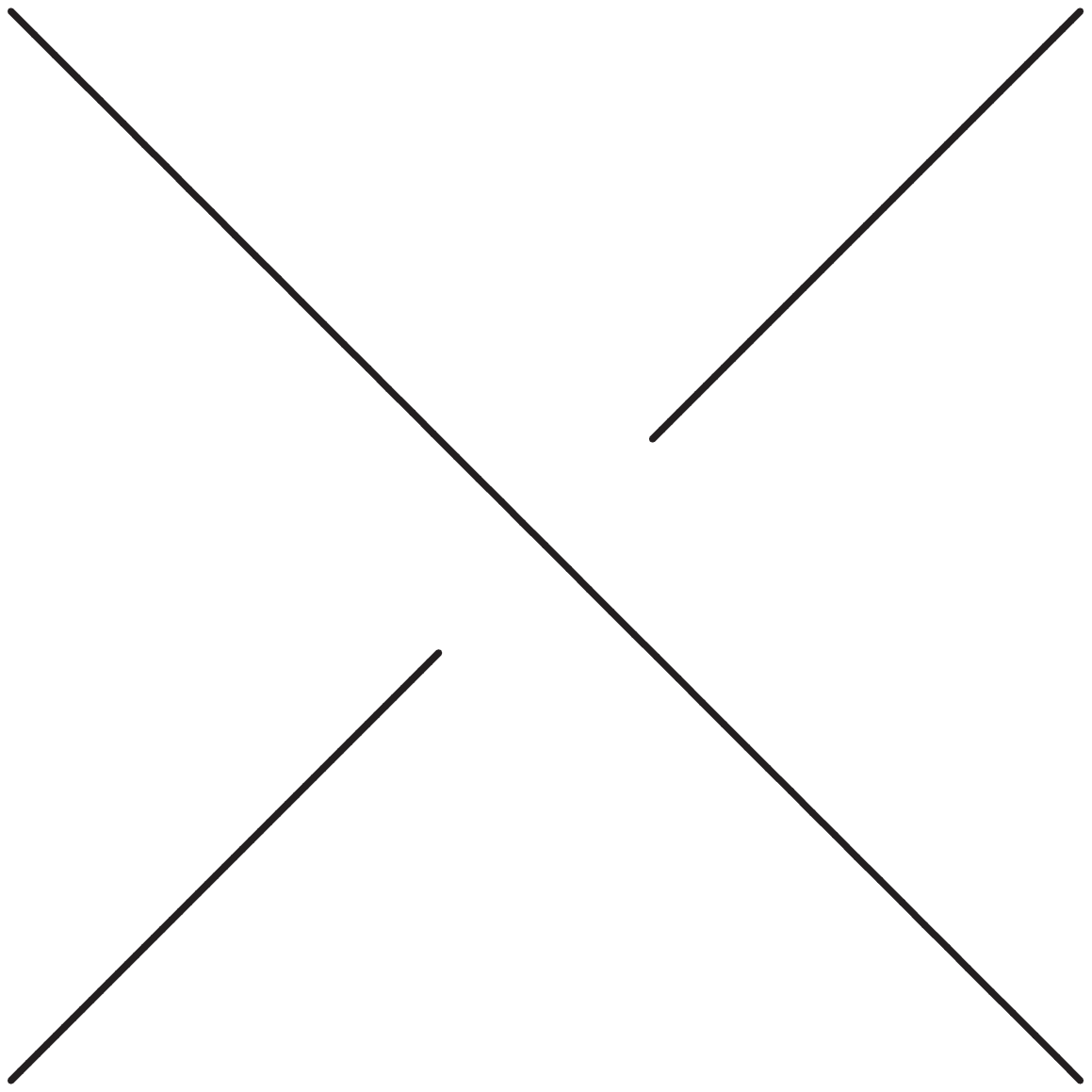}}}
\end{tabular}
\caption{The right twist and the left twist\label{signf}}
\end{center}
\vspace{-10pt}
\end{figure}

To a space curve is associated its diagram, which is given by the
projection on $\RR^2$ and the (under/over) nature
of the crossings. From a diagram one can compute several knot invariants
that may allow to determine the corresponding knot.
It is in general a difficult problem when the minimal number of crossings of the knot is
greater than 16. We will not discuss this question in the present contribution.
\pn
If $a$ and $b$ are coprime integers, then the curve $\cC(a,b,c,\varphi)$
is singular if and only if it has double points.
Let us introduce the polynomials $P_n$ and $Q_n$ defined by
\[
P_n(t,s)= \Frac{T_n(t)-T_n(s)}{t-s},\
Q_n(t,s,\varphi)= \Frac{T_n(t+\varphi)-T_n(s+\varphi)}{t-s}.
\label{PQ} \]
Then, $\cC(a,b,c,\varphi)$ is a knot if and only if
\[
\left \{(s,t), \, P_a(s,t)=P_b(s,t)=Q_c(s,t,\varphi)=0 \right \}
\] is empty.
The projection of the Chebyshev space curve $\cC(a,b,c,\varphi)$
on the $xy$-plane is the plane Chebyshev curve
$$\cC(a,b): x=T_a(t); \, y=T_b(t).$$
The crossing points of $\cC(a,b)$
lie on the $(b-1)$ vertical lines $T'_b(x)=0$ and on the $(a-1)$ horizontal
lines $T'_a(y)=0$. We can represent the knot
$\cC(a,b,c,\varphi)$ by a billiard diagram \citep{KP3} which is a purely
combinatorial object, see for example \citep{CK15}.
As an example, consider the knots $\overline{5}_2 = \cC(4,5,7,0)$,
${5}_2 = \cC(5,6,7,0)$, $\overline{4}_1 = \cC(3,5,7,0)$ in Figure \ref{bt}.
\pn
There are two kinds of crossing: the right twist and the left twist,
see \citep{Mu} and Figure \ref{signf}.
In \citep[Lemma~6]{KPR}, it is shown that
$\cC(a,b)$ has $(a-1)(b-1)/2$
double points $A_{\alpha,\beta}$ corresponding to parameters
($t= 2\cos(\alpha+\beta), s=2\cos(\alpha-\beta)$, $\alpha = \frac{i\pi}a$,
$\beta=\frac{j\pi}b$) and the nature of the crossing over $A_{\alpha,\beta}$
is given by the sign of
\begin{equation}
D(s,t,\varphi) = Q_c(s,t,\varphi) P_{b-a}(s,t)
=(-1)^{i+j + \pent{ib}a + \pent{ja}b} Q_c(s,t, \varphi). \label{D}
\end{equation}

%%%%%%%%%%%%%%%%%%%%%%%%%%%%%%%%%%%%%%%%%%%%
\subsection{The discriminant polynomial}
If $(a,b)=1$, then the algebraic set
$$\cV_{a,b}=\{(s,t) \in \CC^2, \, P_a(s,t)=P_b(s,t)=0\}$$
has exactly $(a-1)(b-1)$ points, and they are real \citep{KP3}.
The leading coefficient of $Q_c$, viewed as a
univariate polynomial in $\varphi$, is equal to $c$ and then
$$\cV_{a,b,c} =
\left \{(s,t,\varphi)\in\CC^3,P_a(s,t)=P_b(s,t)=Q_c(s,t,\varphi)=0\right \}$$
is also a finite set of complex points (see \citet[Prop.~5]{KPR}).
The projection
$$\cZ_{a,b,c} = \{\varphi \in \CC \, \vert \,
\exists (s,t) , \, P_a(s,t)=P_b(s,t)=Q_c(s,t,\varphi)=0\}$$
is also a finite number of points which discriminates the possible
knots: if the interval $(\varphi_1,\varphi_2)$ does not intersect
$\cZ_{a,b,c}$, then
${\cC}(a,b,c,\varphi_1)$ and ${\cC}(a,b,c,\varphi_2)$ represent the same knot,
because the nature of their crossings (in Formula \ref{D}) does not change.
\pn
One can consider $\cZ_{a,b,c}$ as the zero set of $\langle \tilde
R_{a,b,c} \rangle=\langle P_a, P_b,Q_c\rangle\cap \QQ[\varphi]$  or as
the zero set of the characteristic polynomial
$\hat{R}_{a,b,c}$ of $\varphi$ in
$\QQ[s,t,\varphi]\bigm/\langle P_a, P_b ,Q_c\rangle$ which both are
polynomials with rational coefficients that could be computed using
classical elimination tools, see \citet*{KPR}, for example by computing a Gr\"obner basis for any
elimination order with $\varphi<s,t$ or by performing linear algebra
in $\QQ[s,t,\varphi]\bigm/\langle P_a, P_b ,Q_c\rangle$.

The information about the multiplicities of the points of
$\cZ_{a,b,c}$ viewed as roots of $\tilde R_{a,b,c}$ or $\hat{R}_{a,b,c}$
is useless so, in this contribution, we will define
$R_{a,b,c}$ as a polynomial (of degree $\frac{1}{2}(a-1)(b-1)(c-1)$)
with the same roots as $\tilde R_{a,b,c}$ or $\hat{R}_{a,b,c}$ and
we will name it the {\em discriminant polynomial} (for a fixed $(a,b,c)$).

%%%%%%%%%%%%%%%%%%%%%%%%%%%%%%%%%%%%%%%%%%%%%%%%%%%%%%%%%%%%%%%%%
\subsection{Motivations}
In \citep*{BDHZ}, Lissajous knots have been sampled by numerical experiments,
and several knots with relatively small crossing numbers were identified.

Given $a,b,c$ integers, $a,b$ coprime, and $\varphi$
a rational number, our first goal is
\begin{enumerate}
\item[---] decide if $\cC(a,b,c,\varphi)$ is singular;
\item[---] if not, determine its diagram, that is the sign of $D(s,t,\varphi)$
for all $(s,t)$ in ${\cal V}_{a,b}$.
\end{enumerate}
Given $a,b,c$ integers, $a,b$ coprime,
our second goal is to determine all possible diagrams
corresponding to a knot $\cC(a,b,c,\varphi)$.
\begin{enumerate}
\item[---]
compute the discriminant polynomial $R_{a,b,c}(\varphi)$ (or any
polynomial with the same roots as $\tilde R_{a,b,c}$ such that
$\langle \tilde R_{a,b,c} \rangle=\langle  P_a,P_b,Q_c \rangle \cap Q[\varphi]$);
\item[---] compute the real roots $\varphi_1< \cdots <\varphi_s$ of
$R_{a,b,c}(\varphi)$;
\item[---] for an arbitrary set of rational numbers
$r_0<\varphi_1<r_1 < \varphi_2 < \cdots < r_{s-1}<\varphi_s<r_{s}$, compute the $xy$-diagrams
of $\cC(a,b,c,r_i)$.
\end{enumerate}
In \citep{KPR}, the study of Chebyshev knots $\cC(a,b,c,\varphi)$
was restricted
to the case $a \leq 4$, corresponding to {\em two-bridge} knots.
In this case the knots were easily deduced from their diagrams,
by computing the Schubert fraction, see \citep{Mu}.

The discriminant polynomial $R_{a,b,c}$ was directly obtained as a (product of)
resultant(s) with integer coefficients.
The method in \citep{KPR} was essentially based on usual general black-boxes
for solving the zero-dimensional system
$$
\left \{ P_a(s,t)=P_b(s,t)=Q_c(s,t,\varphi)=T-D(s,t,\varphi)=0\right \},
$$
for example by computing a rational univariate representation (RUR, see
\citet{Rouillier1999}) of its zeroes and then compute the
sign of the $T$-coordinate of each real root.

In \citep{KPR} an exhaustive list of minimal parametrization was obtained
for all (but six) two-bridge knots with 10 and fewer crossings,
by enumerating all possible diagrams $\cC(a,b,c,\varphi)$,
for increasing $a<b<c$.
For these six knots one could not find any integer $c$ nor any
rational number $\varphi$, such that $\cC(a,b,c,\varphi)$ was a parametrization.
One of the reason was that the discriminant polynomial
$R_{a,b,c,}$ was too difficult to compute using classical elimination techniques.
\pn
In the present paper, we are not limited to the case $a \leq 4$ anymore and thus, in addition to the description of the
algorithms used, it makes sense to analyse their complexity.

We rather use some remarkable properties of the implicit Chebyshev
curves to factorize the discriminant polynomial %the problem
over the real cyclotomic extension $\QQ[2\cos \pi/n]$, (where $n \leq abc$).
We thus completely change the computational strategy:
one has now to deal with univariate polynomials of low degrees
but with coefficients in some field extensions of high degrees.

We make use of specific properties of Chebyshev
polynomials as well as specific algorithms \citep*{KRT15} working in the Chebyshev basis
instead of the usual monomial basis 
to speed up dramatically the computations.
This new modelization and the related algorithms allow us
to obtain all the classifications of \citep{KPR} in a few minutes,
including the six knots that where not reached.

\subsection{Contents of this paper}

In Section \ref{curves}, we recall some basic properties
of Chebyshev polynomials and then give geometric properties
of Lissajous and Chebyshev curves.
We propose a factorization of  $T_m(x)-T_{m'}(y)$.
The particular case $m'=m$ is used to factorize $R_{a,b,c}$ and
isolate its real roots. Note that this factorisation is also used in \citep{DS12}
in a much more theoretical context.

Section \ref{algo} is devoted to the computation of $R_{a,b,c}$. We
propose a factorization in $\ZZ[\varphi]$ as well as in
$\ZZ[2\cos \pi/n][\varphi]$ with $n=abc$.
We then study two different ways for computing $R_{a,b,c}$:
expressing $R_{a,b,c}$ in $\ZZ[2\cos \pi/n][\varphi]$
as a Chebyshev form or by certified and accurate numerical approximations.
In the first case, we evaluate to $\tcO(n^4)$ bit operations the cost of the
computation, which outperforms the time required by a straightforward
method based on Gr\"obner bases or resultants and in the second case,
we show that the computation requires only $\tcO(n^3)$ bit operations.
All these results are based on results on the cyclotomic extension
$\ZZ[2\cos\pi/n]$ that have been recently published in \citep{KRT15}.

In Section \ref{sec:compcrit}, we focus on the isolation of the real roots of
$R_{a,b,c}$. We first show that the coefficients of $R_{a,b,c}$ are
all bounded in absolute value by $6^N$, with
$N=\frac{1}{2}(a-1)(b-1)(c-1)$ so a direct method using state-of-the-art
algorithms would isolate the real roots in $\tcO(n^3)$ bit operations.
We then propose an ad-hoc method that computes the real roots in
$\tcO(n^2)$ bit operations, thanks to a good separation of the
roots ($>2^{-8n}$). This method does not require to know explicitly the
coefficients of $R_{a,b,c}$.

In Section \ref{sec:diagrams},
we propose some tools for computing all the possible knot diagrams
$\cC(a,b,c,\varphi)$ when $a,b,c \in \ZZ$, $a,b$ coprime, are fixed.
We show that all the possible knot diagrams $\cC(a,b,c,\varphi)$
can be listed in $\tcO(n^2)$ bit operations.
We also show that given $\varphi\in \QQ$ of bitsize $\tau$, it requires
$\tcO(n^2+ n\tau)$ bit operations in order to decide if
$\cC(a,b,c,\varphi)$ is a knot and if so, $\tcO(n^2\tau)$ bit operations
to compute the nature of its crossings.

In the last section, we report the computation we performed
to obtain all two-bridge knots with 10 crossings and fewer.

%%%%%%%%%%%%%%%%%%%%%%%%%%%%%%%%%%%%%%%%%%%%%%%%%%%%%%%%%%%%%%%%%%%%%%
\section{Chebyshev and Lissajous curves}\label{curves}
In this section, we show that the implicit Chebyshev curve $T_b(x)=T_a(y)$
factorizes in Lissajous curves, which allows us to give explicit factorizations for
$P_a,P_b$ and $Q_c$ that will intensively be used in the sequel.
\pn
Chebyshev polynomials and their algebraic properties play a
central role. The monic Chebyshev polynomials of the {\em first kind},
also called Dickson polynomials,
are defined by the second-order linear recurrence
\[ T_0 = 2, \, T_1 = t, \, T_{n+1} = tT_n - T_{n-1}.\label{rl1}\]
They satisfy the identity $T_n(2\cos\theta) = 2\cos n\theta$, and then
$T_n \circ T_m = T_{nm}$.
The monic Chebyshev polynomials of the {\em second kind}
satisfy
$V_n(2\cos\theta) = \Frac{\sin n\theta}{\sin\theta}$ and
$T'_n = n V_n$.
Both $T_n(t)$ and $V_n(t)$ belong to $\ZZ[t]$ and we have
$$T_n = \Prod_{k=0}^{n-1} (t-2\cos\frac{(2k+1)\pi}{2n}), \quad
V_n = \Prod_{k=1}^{n-1} (t-2\cos\frac{k\pi}n).$$
\pn
The following classical properties will be useful in this section:
\begin{lem}\label{lem:cp} Let $T_n$ be the monic Chebyshev polynomial of 
the first kind.
\bi
\item If $T'_n(t)=0$ then $T_n(t)=\pm 2$,
if $T_n(t)=\pm 2$ then $T'_n(t)=0$ or  $t = \pm 2$.
\item $T_n(t)=y$ has $n$ real solutions if and only if $\abs y <2$.
$T_n(t)=2$ has $\pent n2$ real solutions.
$T_n(t)=-2$ has $\pent{n-1}2$ real solutions.
\ei
\end{lem}
\Pf
From $T'_n = n V_n$, we deduce that $t\mapsto T_n(t)$ is monotonic when
$\abs t \geq 2\cos\frac{\pi}n$ and
that $T_n$ has $n-1$ local extrema for $t_k = 2\cos\frac{k\pi}n$ where
$T_n(t_k)=2(-1)^k$.
\EPf
\pn
The following proposition gives a unified definition of Lissajous and
Chebyshev curves:
\begin{prop} \label{equa}
Let $a,b$ be coprime integers ($a$ odd) and $\varphi \in \RR$.
The parametric curve
$$\cC: x=2\cos( at), \, y = 2\cos(bt+\varphi), \, t \in \CC,$$
admits the equation $C_{a,b,\varphi} = 0$ where
\[
C_{a,b,\varphi} = T_b(x)^2 + T_a(y)^2 - 2 \cos(a\varphi)T_b(x)T_a(y) - 4\sin^2(a\varphi).\label{eqE}
\]
\begin{enumerate}
\item If $a\varphi \not = k \pi$, then $C_{a,b,\varphi}$ is irreducible.
$\cC$ is called a Lissajous curve.
Its real part is one-to-one parametrized for $t \in [0,2\pi]$.
\item If $a\varphi = k \pi$, then $C_{a,b,\varphi} = \bigl (T_b(x)-(-1)^k T_a(y) \bigr)^2$.
$\cC$ is called a Chebyshev curve. It can be one-to-one parametrized
by $x= T_a(t), \, y=(-1)^k T_b(t)$.
\end{enumerate}
\end{prop}
\Pf
Let $(x,y) \in \cC$.
We have $T_b(x)=2\cos( ab\, t), \, T_a(y) = 2\cos(ab\,t + a\varphi)$.
Let $\lambda = a\varphi, \, \theta=abt$. We get
$T_a(y) = 2\cos(\theta+\lambda)$ so $4(1-\cos^2 \theta)\sin^2\lambda = (2\cos\theta \cos\lambda - T_a(y))^2$, that is
$(4-T_b^2(x)) \sin^2\lambda = (T_b(x)\cos\lambda - T_a(y))^2,$ and we deduce our Equation (\ref{eqE}).
\pn
Conversely, suppose that $(x,y)$ satisfies (\ref{eqE}). Let
$x=2\cos(at)$ where $t \in \CC$. We also have $x=2\cos
a(t+\frac{2k\pi}a)$ and $T_b(x)=2\cos\theta$. $A=T_a(y)$ is solution of the second-degree equation
$$A^2 - 2\, A \, \cos(a\varphi)\cos\theta - 4\sin^2(a\varphi)=0.$$
Consequently, we get %$A= \cos(\theta \pm a\varphi)$ and
$T_a(y) = 2\cos(\theta \pm a\varphi) = T_a(\cos(\pm bt + \varphi))$.
We deduce that $y = 2\cos(\pm bt + \varphi + \frac{2h\pi}a)$, $h \in \ZZ$.
Changing $t$ by $-t$, we can suppose that
$$ x=2\cos at, \, y = 2\cos(bt+\varphi +\frac{2h\pi}{a}).$$
By choosing $k$ such that $kb+h\equiv 0 \Mod a$, we get $
x=2\cos at', \, y = 2\cos(bt'+\varphi),$ where $t'=t+\frac{2k\pi}a$.
\pn
If $a\varphi \not \equiv 0 \Mod \pi$. Suppose that Equation (\ref{eqE}) factors in
$P(x,y) Q(x,y)$. We can suppose, for analyticity reasons, that $P(2\cos (at) , 2\cos (bt+\varphi))=0$,
for $t\in \CC$. The curve $\cC$ intersects the line $y=0$ in $2b$ distinct points
so $\deg_x P \geq 2b$. Similarly, $\deg_y P \geq 2a$ so that $Q$ is a constant
which proves that the equation is irreducible.
\pn If $\cos a\varphi = (-1)^k$, the equation becomes $T_b(x)-(-1)^k T_a(y)=0$.
In this case the curve admits the announced parametrization, see \citep{Fi} and \citep{KP3} for more details.
\EPf
%%%%%%%%%%%%%%%%%%%%%%%%%%%%%%%%%%%%%%%%%%%%%%%%%%%%%%%%%%%%%%%%
\begin{defn}
Let $E_{\mu}(x,y) = x^2+y^2- 2 \cos(\mu)xy - 4\sin^2(\mu)$ when
$\mu \not\equiv 0 \Mod{\pi}$ and $E_0 = x-y$, $E_{\pi} = x+y$.
Equation (\ref{eqE}) is equivalent to $E_{a\varphi} (T_b(x),T_a(y))=0$.
\end{defn}
%\pn
\begin{rem}\label{rem:inscribed}
If $a=b=1$, we obtain the Lissajous ellipses.
They are the first curves studied by Lissajous \citep{Li}.
Let $\mu \not \equiv 0 \Mod{\pi}.$ The curve
$E_{\mu}(x,y)=0$ is an ellipse $\cE_{\mu}$
inscribed in the square $[-2,2]^2$. It admits the
parametrization $x=2\cos t, \, y=2\cos (t +\mu)$.
This shows that the real part of the curve $\cC$ (Equation (\ref{eqE})) is
inscribed in the square $[-2,2]^2$.
\end{rem}
\pn
\begin{figure}[!th]
\begin{center}
\begin{tabular}{ccc}
\epsfig{file=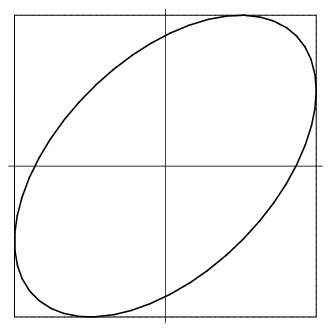,height=3.5cm,width=3.5cm}&
\epsfig{file=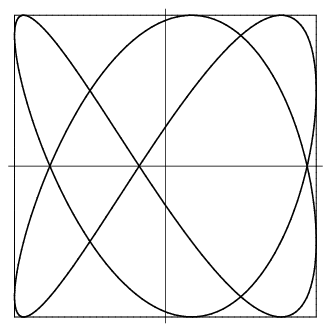,height=3.5cm,width=3.5cm}&
\epsfig{file=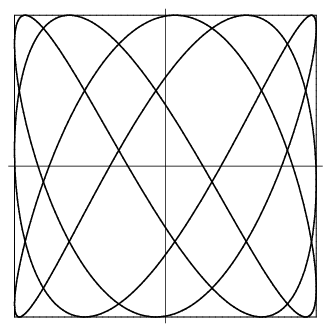,height=3.5cm,width=3.5cm}\\
{\small $a=b=1, \varphi=\pi/3$}&{\small $a=2,b=3,\varphi=\pi/3$}&
{\small $a=3,b=5,\varphi=2\pi/15$}
\end{tabular}
\caption{Lissajous curves $x=2\cos at, y= 2 \cos (bt + \varphi)$}
\label{fig:lc}
\end{center}
\end{figure}
\pn
Using Proposition \ref{equa},
we recover the following classical result.
\begin{cor}
The Lissajous curve $x=2 \cos(at), \, y=2 \cos(bt+\varphi)$,
($a\varphi\not \equiv 0 \Mod \pi$) has $2ab - a - b$ singular points
which are real double points.
\end{cor}
\Pf
The singular points of $\cC$ satisfy Equation (\ref{eqE}) and the system
$$
\left \{
\begin{array}{rcl}
T'_b(x) (T_b(x) - T_a(y) \cos a\varphi)=0, \\
T'_a(y) (T_a(y) - T_b(x) \cos a\varphi)=0.
\end{array}
\right .
$$
Suppose that $T'_b(x)=T'_a(y)=0$, then $T^2_a(y) = T^2_b(x)=4$ from Lemma
\ref{lem:cp}. Equation (\ref{eqE}) is not satisfied
since $\cos a \varphi \not = \pm 1$.
Suppose that $T_b(x) - T_a(y) \cos a\varphi=T_a(y) - T_b(x) \cos a\varphi=0$,
then $T_b(x)=T_a(x)=0$ and Equation \ref{eqE} is not satisfied.
We thus have either $T'_b(x)=0$ and $T_a(y) - T_b(x) \cos a\varphi=0$ that
gives $(b-1)\times a$ real points because of the classical properties of
Chebyshev polynomials, or $T'_a(y)=0$ and $T_b(x) - T_a(y) \cos a\varphi=0$
that gives $b\times (a-1)$ real double points.
\EPf
\begin{rem}
The study of the double points of Lissajous curves is classical
(see \citet{BHJS} for  their parameter values). The study of the double
points of Chebyshev curves is simpler \citep{KP3}.
\end{rem}
\begin{cor}
The affine implicit curve $T_n(x)=T_m(y)$ has
$\pent{n-1}2\pent{m-1}2 + \pent{n}2\pent{m}2$ singular points that
are real double points.
\end{cor}
\Pf
The singular points satisfy either $T_n(x)=T_m(y)=2$ or $T_n(x)=T_m(y)=-2$
and we conclude using Lemma \ref{lem:cp}.
\EPf
\begin{thm}{\bf Factorization of $\mathbf{T_n(x)-T_n(y)}$.}\label{thm:fell}
We have
\[
\Frac{T_n(t)-T_n(s)}{t-s} = \Prod_{k=1}^{\pent{n}{2}} E_{\frac{2k\pi}n}(s,t).
\label{eq:fell}
\]
\end{thm}
\Pf Following \citet{Tr15}, let
$(t,s) \in \cE_{2k \frac{\pi}n}$, then
$t = \cos \rho$, $s=\cos (\rho + 2k \frac{\pi}n)$ and
$T_n(t)=T_n(s)$.
Since the polynomials $E_{2k \frac{\pi}n}$ are distinct and irreducible, we obtain 
$T_n(t)-T_n(s) = \Prod_{k=0}^{\pent{n}{2}} E_{\frac{2k\pi}n}(s,t)$.
\EPf
The curve  $\Frac{T_n(t)-T_n(s)}{t-s}=0$ has $\pent n2$ irreducible components.
It is a union of ellipses $\cE_{\frac{2k\pi}n}$ and at most one line ($x+y=0$).
Note that $\cE_{\frac{2k\pi}n}$ and $\cE_{\frac{2l\pi}m}$
intersect at the point
$(t,s) = (2\cos(\frac{k\pi}{n}+\frac{l\pi}{m}),
2\cos(\frac{k\pi}{n}-\frac{l\pi}{m}))$
and its reflections with respect to the lines $s=-t$ and $s=t$.
We recover the parametrization
of the double points of
$x=T_a(t), \, y=T_b(t)$ that will be very useful for the description of Chebyshev space curves:
\begin{prop}\citep{KP3,KPR}\label{dp}
Let $a$ and $b$ are nonnegative coprime integers, a being odd. Let
the Chebyshev curve $\cC$ be defined by  $ x= T_a(t), \  y=T_b(t).$
The pairs $(t,s)$ giving a crossing point are
$$t=2\cos (\frac{j\pi}b+\frac{i\pi}a), \ s=2\cos(\frac{j\pi}b-\frac{i\pi}a)$$
where $1\leq i \leq \frac 12 (a-1)$, $1\leq j \leq b-1$.
\end{prop}
\pn
\begin{figure}[!th]
\begin{center}
\begin{tabular}{ccc}
\epsfig{file=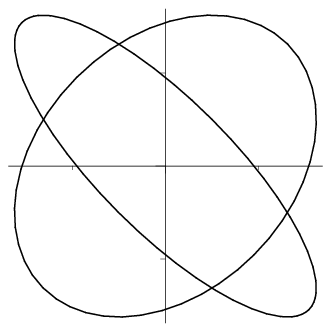,height=3.5cm,width=3.5cm}&{\scalebox{1}{\includegraphics{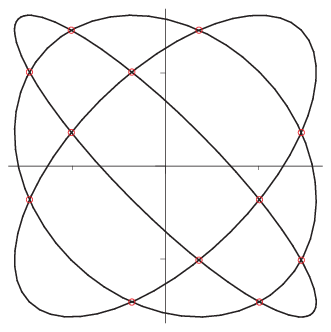}}}&
\epsfig{file=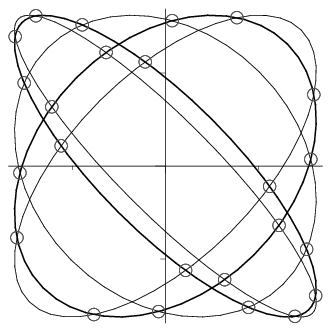,height=3.5cm,width=3.5cm}\\
{\small $\Frac{T_5(t)-T_5(s)}{t-s}=0$}&{\small $\Frac{T_7(t)-T_7(s)}{t-s}=0$}&
{\small
$\left \{
\begin{array}{c}
\Frac{T_7(t)-T_7(s)}{t-s}=0\\
\Frac{T_5(t)-T_5(s)}{t-s}=0
\end{array}\right .$}
\end{tabular}
\caption{Double point in the parameter space}
\label{T7}
\vspace{-10pt}
\end{center}
\end{figure}
\pn
We thus deduce
%%%%%%%%%%%%%%%%%%%%%%%%%%%%%%%%%%%%%%%%%%%%%%%%%%%%%%%%%%%%%%%%%%%%%%%
\begin{cor}{\bf Factorization of $\mathbf{T_n(x)-T_m(y)}$.}\label{fact}\\%
Let $m=ad$, $n=bd$, $(a,b)=1$ and $a$ odd. We have the factorization
$$T_n(x)-T_m(y) = \Prod_{k=0}^{\pent d2} C_k(x,y)$$
where $C_k(x,y) = E_{\frac{2k\pi}d}(T_b(x),T_a(y))$.
\end{cor}
\Pf
We get $T_n(x)-T_m(y) = T_d(T_b(x)) - T_d(T_a(y))$ and we conclude using
Theorem \ref{thm:fell}.
\EPf
\begin{cor}
Let $d=\gcd(n,m)$. The curve $T_n(x)=T_m(y)$ has $\pent d2 +1$ components,
$\pent{d-1}2$ of them are Lissajous curves.
\end{cor}
\begin{figure}[!th]
\begin{center}
\begin{tabular}{ccc}
\epsfig{file=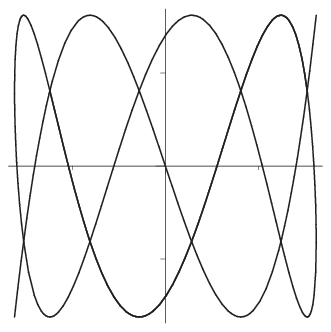,height=3.5cm,width=3.5cm}&
\epsfig{file=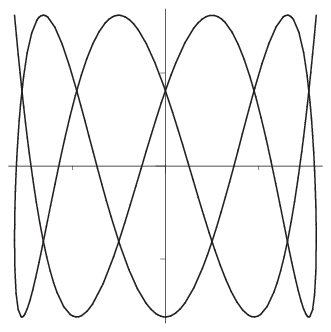,height=3.5cm,width=3.5cm}&
\epsfig{file=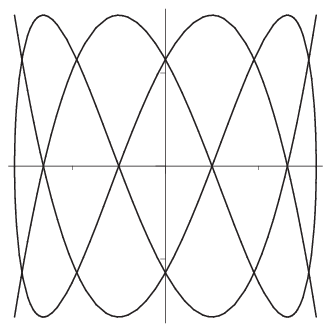,height=3.5cm,width=3.5cm}\\
{\small $T_{9}(x)=T_3(y)$}&{\small $T_{10}(x)=T_3(y)$}&
{\small $T_{10}(x)=T_4(y)$}
\end{tabular}
\caption{Implicit Chebyshev curves}
\label{Td}
\end{center}
\end{figure}

\def\cT{{\cal T}}
\def\Deltanul{{\Delta_{\alpha,\beta,\gamma}}}
\def\Pnul{{P_{\alpha,\beta,\gamma}}}
\def\Pnulhat{{\hat{P}_{\alpha,\beta,\gamma}}}
\def\Pijk{{P_{\tfrac{i\pi}{a},\tfrac{j\pi}{b},\tfrac{k\pi}{c}}}}
\def\Ptijk{{\widetilde P_{\tfrac{i\pi}{a},\tfrac{j\pi}{b},\tfrac{k\pi}{c}}}}
\def\cb{{C_{\cal B}}}
\def\cbi#1{C_{\mathcal B}^{(#1)}}
\def\Ll{{\cal L}}
\def\qc{{Q_{c,\alpha,\beta}}}
\def\taut{\tau}
%%%%%%%%%%%%%%%%%%%%%%%%%%%%%%%%%%%%%%%%%%%%%%%%%%%%%%%%%%%%%%
\section{Computing the discriminant polynomial $R_{a,b,c}$}\label{algo}
In this section, we will study several methods to compute  efficiently
a polynomial $R_{a,b,c}\in \QQ[\varphi]$ whose roots are
$$\cZ_{a,b,c} = \{\varphi \in \CC \vert \,
\exists (s,t), \, P_a(s,t)=P_b(s,t)=Q_c(s,t,\varphi)=0\}.$$

We will propose a formal computation of $R_{a,b,c}$ using results
from \citet{KRT15}
on fast operations on Chebyshev forms with a bit complexity in $\tcO(n^4)$ bit
operations (with $n=abc$) and also a method using
approximate computations (but still providing the exact result)
with a bit complexity in $\tcO(n^3)$ operations.

One might use for $R_{a,b,c}$ the generator of the principal ideal
$\langle P_a,P_b,Q_c \rangle \cap \QQ[\varphi]$ which can
thus be obtained from any Gr\"obner
basis $\langle P_a,P_b,Q_c  \rangle$ for any
monomial
%ordering
order such that $\varphi < s,t$.

Such a straightforward method could be optimized using the structure
of the system: $P_a,P_b\in \QQ[s,t]$, $Q_c\in \QQ[s,t,\varphi]$ and,
moreover, the fact that the leading coefficients of $Q_c$ with
respect to $\varphi$ belongs to $\ZZ$, see Formula (\ref{Qc}). Then, one can first compute
a Gr\"obner basis $G_{a,b}$ of $\langle P_a,P_b \rangle$ for any order
$<_{a,b}$ and then obtain, without computation, a Gr\"obner basis $G_{a,b,c}$ of
$\langle P_a,P_b,Q_c \rangle$ for any order
compatible with $<_{ab}$ and such that $s,t<\varphi$,
by just adding $Q_c$ to $G_{a,b}$. Even if the computation time for getting
$G_{a,b}$ could be neglected in practice since $a,b<<c$, even if $G_{a,b,c}$
could be easily obtained from $G_{a,b}$, computing $R_{a,b,c}$ still requires to
compute the minimal polynomial of $\varphi$ in
$\QQ[s,t,\varphi] \bigm/ \langle P_a,P_b,Q_c \rangle$
with almost no hope to reach the announced binary complexities.

\subsection{The discriminant polynomial $R_{a,b,c}$ }

As specified %precised
in the introduction, the information on
multiplicities of the roots of $R_{a,b,c}$ is useless so that the following
proposition gives an admissible definition for $R_{a,b,c}$:

\begin{prop}\label{prop:rabc1}
Let $a$, $b$ be coprime integers, $a$ odd, and let $c$ be an integer.
Let us consider the polynomial
$$R_{a,b,c}(\varphi) =
\Prod_{i=1}^{\frac{a-1}2}\Prod_{j=1}^{b-1}
Q_c(2\cos(\frac{i\pi}a+\frac{j\pi}b),2\cos(\frac{i\pi}a-\frac{j\pi}b),\varphi).$$
Then $ R_{a,b,c}\in \ZZ[\varphi]$ and $\cC(a,b,c,\varphi)$
is singular if and only if $ R_{a,b,c}=0$.
\end{prop}
\Pf
The curve $\cC(a,b,c,\varphi)$ is singular if and only if
it admits double points.
This condition is equivalent to have
$t=2\cos (\frac{j\pi}b+\frac{ i\pi}a)$ and
$s=2\cos (\frac{j\pi}b-\frac{ i\pi}a)$ and $Q_c(s,t,\varphi)=0$, for some
$1\leq i \leq \frac{a-1}2$ and $1\leq j \leq b-1$, from Proposition \ref{dp}.
We thus deduce that $\cC(a,b,c,\varphi)$ is singular if and only if $\varphi$
is a root of $$ R_{a,b,c}(\varphi) =
\Prod_{i=1}^{\frac{a-1}2}\Prod_{j=1}^{b-1}
Q_c(2\cos(\frac{i\pi}a+\frac{j\pi}b),2\cos(\frac{i\pi}a-\frac{j\pi}b),\varphi).
$$
$Q_c(s,t,\varphi)$ is a symmetrical polynomial of $\ZZ[\varphi][t,s]$.
Let $\alpha_i = \frac{i\pi}a$, $\beta_j = \frac{j\pi}b$ and
$s=2\cos(\alpha_i+\beta_j)$, $t=2\cos(\alpha_i-\beta_j)$.
From $s+t = 4 \cos\alpha_i \cos\beta_j$ and
$st= 2\cos 2\alpha_i +2\cos 2\beta_j$, we deduce that
$Q_c(s,t,\varphi)$  belongs to $\ZZ[\varphi,2 \cos\alpha_i] [2\cos\beta_j]$.
\[\label{eq:Ri}
R_i = \Prod_{j=1}^{b-1} Q_c(2\cos(\alpha_i+\beta_j),\cos(2\alpha_i-\beta_j),\varphi)
\]
belongs to $\ZZ[\varphi,2\cos\alpha_i]$ because
the roots of $V_b \in \ZZ[t]$ are the $2\cos\beta_j$, $j=1,\ldots,b-1$.
From $Q_c(-s,-t,-\varphi)=(-1)^{c-1} Q_c(s,t,-\varphi)$ we deduce that
$\Prod_{i=1}^{{a-1}}R_i(\varphi) = \pm \Prod_{i=1}^{\frac{a-1}2} R_i(-\varphi)R_i(\varphi)
\in \ZZ[\varphi]$.
We thus have $R^2_{a,b,c} \in \ZZ[\varphi]$ and so it is for $R_{a,b,c}$.
\EPf
\begin{cor}\label{cor:2res}
Let $R_c (2\cos \alpha, 2 \cos \beta,\varphi) =
Q_c(2\cos(\alpha+\beta),\cos(2\alpha-\beta),\varphi)$.
Then we have $ R^2_{a,b,c} = \Res_u(\Res_v(R_c(u,v,\varphi),V_a),V_b)$.
\end{cor}
\Pf
In the proof of  Proposition \ref{prop:rabc1}, we have
$R_i = \Res_v(R_c(2\cos\alpha_i,v,\varphi),V_b(\varphi))$ in
Formula (\ref{eq:Ri}) and then
$\Res_u(\Res_v(R_c(u,v,\varphi),V_b),V_a) = \Prod_{i=1}^{a-1} R_i(\varphi) =
 R_{a,b,c}^2.$
\EPf
\pn
{\bf Example.}
When $a=3$, $b=4$, $c=5$, we find that
$$R_{a,b,c} =
 \left( 5{\varphi}^{4}+15{\varphi}^{2}-1 \right) \cdot
\left( 25{\varphi}^{8}-50{\varphi}^{6}+35{\varphi}^{4}-20{\varphi}^{2}+1 \right).$$
There are exactly 6 critical values that are symmetrical about the origin.
\begin{figure}[!th]
\begin{center}
\begin{tabular}{ccccc}
\epsfig{file=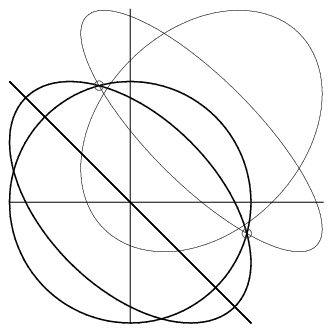,height=4cm,width=4cm}&&
\epsfig{file=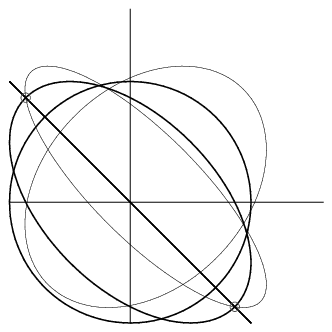,height=4cm,width=4cm}&&
\epsfig{file=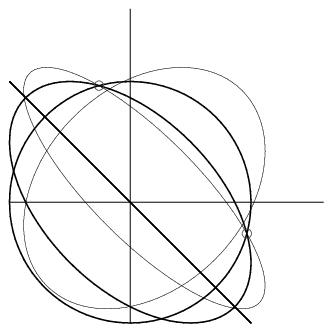,height=4cm,width=4cm}\\
{\small $\varphi=.590$}&&{\small $\varphi=.128$}&&
{\small $\varphi=.117$}\\
\end{tabular}
\caption{$P_3=0, P_4=0, Q_5=0$}
\label{T345}
\end{center}
\end{figure}
For these values of $\varphi$, the curve $Q_5(s,t,\varphi)=0$, which is
translated from the curve $P_5(s,t)=0$ by the vector
$(\varphi,\varphi)$, meets the points $\{P_3=0, P_4=0\}$ (see Figure \ref{T345}).
%%%%%%%%%%%%%%%%%%%%%%%%%%%%%%%%%%%%%%%%%%%%%%%%%%%%%%%%%%%%%%%%%%%

%%%%%%%%%%%%%%%%%%%%%%%%%%%%%%%%%%%%%%%%%%%%%%%%%%%%%%%%%%%%%%%%%%%%%%%%%%%%%%
\subsection{Factorizing $R_{a,b,c}$ into low-degree polynomials}
Formula (\ref{eq:fell}) will give us an
explicit formula for the polynomial $R_{a,b,c}$ as a product of
polynomials of degree 1 or 2 with coefficients in
$\ZZ[2\cos\frac\pi{a},2\cos\frac\pi{b},2\cos\frac\pi{c}]$. The
interest of having such an expression is to make possible the use of
efficient tools for evaluating trigonometric expressions, such as
those proposed in \citep{KRT15}.

Let us introduce the following polynomials:
\begin{defn}
We set $P_{\alpha,\beta,\frac\pi{2}}(\varphi) =\varphi+2\cos\alpha\cos\beta$ and
$$P_{\alpha,\beta,\gamma}(\varphi) = \varphi^2+4 \varphi \cos\alpha \cos\beta+
4 \Frac{(\cos^2\alpha-\cos^2\gamma)(\cos^2\beta-\cos^2\gamma)}{\sin^2\gamma},$$
for $\gamma \not = \frac{\pi}2$.
\end{defn}
When $\gamma \not = \frac{\pi}2$, we have
$4 \sin^2 \gamma P_{\alpha,\beta,\gamma} =
E_{2\gamma}(2\cos(\alpha+\beta)+\varphi,2 \cos(\alpha-\beta)+\varphi)$.
When $\gamma=\frac{\pi}2$, we have $2 \sin \gamma  P_{\alpha,\beta,\gamma}
= E_{\pi}$. Then, using Equation (\ref{eq:fell}) and
$\Prod_{k=1}^{c-1} 2\sin\frac{k\pi}c=c$, we deduce that
\begin{equation}
Q_c(2\cos(\alpha+\beta), 2 \cos(\alpha-\beta),\varphi)
=c \Prod_{k=1}^{\pent c2} P_{\alpha,\beta,\frac{k\pi}{c}}(\varphi),
\label{Qc}
\end{equation}
and we get the factorization of the polynomial $R_{a,b,c}$:
\begin{prop}\label{prop:rabc2}
Let $a,b$ be nonnegative coprime integers, $a$ odd, and $c$ be an integer.
Then
\begin{equation}
R_{a,b,c}(\varphi) = c^{(a-1)(b-1)/2}
\Prod_{i=1}^{\frac{a-1}2}\Prod_{j=1}^{b-1}\Prod_{k=1}^{\pent c2}
P_{\frac{i\pi}{a},\frac{j\pi}{b},\frac{k\pi}{c}}(\varphi).
\label{eq:rabc2}
\end{equation}
\end{prop}
We have written $R_{a,b,c}$ as the product of second or first-degree
polynomials $P_{\alpha,\beta,\gamma}$ in
$\ZZ[2\cos\frac\pi{a},2\cos\frac\pi{b},2\cos\frac\pi{c}][\varphi]$.
\begin{rem}
\label{rem:rabc12}
There are two cases to consider in Formula (\ref{eq:rabc2}).
If $c$ is odd then $R_{a,b,c}$ appears as the product
\[
\label{eq:rabc1}
R^{(1)}_{a,b,c} (\varphi)=
\Prod_{i=1}^{\frac{a-1}2}\Prod_{j=1}^{b-1}\Prod_{k=1}^{\frac{c-1}2}
4 \sin^2 \frac{k\pi}{c} P_{\frac{i\pi}{a},\frac{j\pi}{b},\frac{k\pi}{c}}(\varphi).
\]
If $c$ is even we have to multiply the previous product by
\[
\label{eq:rabc0}
R^{(0)}_{a,b,c} (\varphi)=
\Prod_{i=1}^{\frac{a-1}2}\Prod_{j=1}^{b-1} (2 \varphi + 4 \cos \alpha\cos\beta).
\]
\end{rem}
%%%%%%%%%%%%%%%%%%%%%%%%%%%%%%%%%%%%%%%%%%%%%%%%%%%%%%%%%%%%%%%%%%%%
\subsection{Computing $R_{a,b,c}$ using Chebyshev polynomials}

In this section, we will use the algorithms
for evaluating and using trigonometric
expressions in the form $F=\sum_{k=0}^df_k\cos k\frac{\pi}{n}$,
$f_k\in\ZZ$, that are developed in \citep{KRT15}.

The Chebyshev basis $(1, T_i, i\geq 1)$ is particularly adapted for our
computations. We say that $f=f_0 + \sum_{i=1}^{d} f_i T_i$, $f_i \in \ZZ$, is a
{\em Chebyshev form}.
\begin{defn}
Let $f=f_0 + \sum_{i=1}^{d} f_i T_i$ be a Chebyshev form. We denote
by $\taut(f)$ the maximum bitsize of its coefficients. We denote by
$\norm{f}_T$ the norm $\abs{f_0} + 2 \sum_{i=1}^d \abs{f_i}$.
\end{defn}
The cyclotomic extension $\QQ[2\cos \frac{\pi}n]$ is $\QQ[x]/(M_n)$ where
$M_n$ is the minimal polynomial in $\ZZ[x]$ of $2\cos \frac{\pi}n$.
In \citep{KRT15}, it is shown that $M_n$ is monic of degree $\frac 12 \phi(2n)$, where $\phi: \NN^* \to \NN$ is the Euler totient function.
$M_n$ can be computed in the Chebyshev basis
in $\tcO(n)$ arithmetic operations or $\tcO(n^2)$ bit operations
and $\taut(M_n) = \cO(n)$, see \citep[Prop.~16]{KRT15}.

%%%%%%%%%%%%%%%%%%%%%%%%%%%%%%%%%%%%%%%%%%%%%%%%%%%%%%%%%
\begin{lem}\citep{KRT15}\label{lem:fgmodm}
Let $f=f_0 + \sum_{i=1}^{n-1} f_i T_i$ and
$g=g_0 + \sum_{i=1}^{n-1} g_i T_i$ be Chebyshev forms with
$\taut(f), \taut(g)\leq \tau$.
Then one can compute $h = h_0 + \sum_{i=1}^{n-1} h_i T_i$ where
$h \equiv f\cdot g \Mod{M_n}$ in $\tcO(n\tau)$ bit operations
and $\taut(h) \leq \taut (f) + \taut (g) + \log_2 n + 1$.
\end{lem}
We thus deduce
\begin{cor}\label{cor:prod}
Let $P = \sum_{i=0}^D p_i(2\cos\frac{\pi}n) \varphi^i$ and
$Q = \sum_{i=0}^{d} q_i(2\cos\frac{\pi}n) \varphi^i$ be two polynomials in
$\ZZ[2\cos\frac{\pi}{n}][\varphi]$. Suppose that the Chebyshev forms
$p_i = \sum_{j=0}^{n-1} p_{i,j} T_j$ and
$q_i = \sum_{j=0}^{n-1} q_{i,j} T_j$
satisfy $\taut(p_i) \leq \tau$ and $\taut(q_i) \leq \tau'\leq \tau$.

Then we have $P \cdot Q = \sum_{i=0}^{D+d} h_i(2\cos\frac{\pi}n) \varphi^i$
where $h_i = \sum_{j=0}^{n-1} h_{i,j} T_j$ satisfy
$\taut(h_i)\leq \tau + \tau' + \log_2 n +\log_2 d $.
One can compute all the Chebyshev forms $[\cT]h_i$ in
$\cO(dD)$ operations in $\ZZ[2\cos\frac{\pi}n]$ that is
$\tcO(dD n\tau)$ binary operations.
\end{cor}
\Pf We get $P\cdot Q = \sum_{i=0}^{d+D} h_i \varphi^i$ where
$h_i = \sum_{j=0}^d p_{i-j} q_j$.
Each $p_{i-j} q_j$ may be computed in $\tcO(n\tau)$ binary operations and
therefore all the coefficients $h_i$ may be computed in $\tcO(d(d+D)n\tau)$
binary operations, using Lemma \ref{lem:fgmodm}.
Furthermore $\tau(h_i) \leq \tau+\tau' + \log_2 n + \log d$, using Lemma
\ref{lem:fgmodm}.
\EPf
\pn
We deduce
\begin{cor}\label{cor:prod2}
Let $P_i = \sum_{j=0}^d p_{i,j}(2\cos\frac{\pi}n) \varphi^j$ be polynomials
in $\ZZ[2\cos\frac{\pi}{n}][\varphi]$ with $\taut(p_{i,j}) \leq \tau$.
Then we have $P = \Prod_{i=1}^k P_i = \sum_{i=0}^{dk} h_i(2\cos\frac{\pi}n) \varphi^j$
where $h_i = \sum_{j=0}^{n-1} h_{i,j} T_j$ satisfy
$\taut(h_i) \leq k\tau + k \log_2 nd$.
One can compute $P$ in $\tcO(d^2k^3n\tau)$ binary operations.
\end{cor}
\Pf
Let $Q_i = P_1 \cdots P_i$. Then we have, from Corollary \ref{cor:prod},
$\taut(Q_{i+1}) \leq \taut(Q_i) + \tau + \log_2 nd \leq (i+1) (\tau+\log_2 nd)$.
One computes $Q_{i+1}$ from $Q_i$ in $\tcO(\taut(Q_i)d(id+d)n\tau) =
\tcO(i^2 d^2 n \tau)$ binary operations, using Corollary \ref{cor:prod}.
At the end we get $Q_k$ in $\tcO(k^3 d^2 n \tau)$ binary operations.
\EPf
\pn
We then compute $R_{a,b,c}$ in
$\ZZ[2\cos \frac{\pi}a,2\cos \frac{\pi}b,2\cos \frac{\pi}c][\varphi]
\subset \ZZ[2\cos \frac{\pi}n][\varphi]$, using
Formulas (\ref{eq:rabc1}) and (\ref{eq:rabc2}) and Corollary \ref{cor:prod2}.
%%%%%%%%%%%%%%%%%%%%%%%%%%%%%%%%%%%%%%%%%%%%%%%%%%%%%%%%%%%%%%%%%%
\begin{prop}\label{prop:rabc-comp}
Let $a$ and $b$ be coprime integers, $a$ odd, and $c$ an integer.
One can compute $R_{a,b,c}$ as an element of
$\ZZ[2\cos \frac{\pi}n][\varphi]$ in $\cO(n^4)$ binary operations.
\end{prop}
\Pf
We want to compute the product of the polynomials
$
4 \sin^2 \frac{k\pi}c \cdot
P_{\frac{i\pi}a,\frac{j\pi}b,\frac{k\pi}c} (\varphi)
$
in Formula (\ref{eq:rabc1}).
We write, for $k\frac{\pi}{c} \not = \frac{\pi}2$:
$$
4 \sin^2 \frac{k\pi}c \cdot
P_{\frac{i\pi}a,\frac{j\pi}b,\frac{k\pi}c} (\varphi)
= \Frac 14 \Bigl ( f_2 (2\cos\frac{\pi}n)\varphi^2 + f_1(2\cos\frac{\pi}n) \varphi
+ f_0(2\cos\frac{\pi}n) \Bigr )
$$
where
\begin{equation}\label{eq:cf2}
\begin{array}{rcl}
f_2&=&2-T_{{2\,kab}},\\
f_1&=&2\,T_{{cja-cib}}+2\,T_{{cja+cib}}-T_{{2\,kab+cja-cib}}
-T_{{2\,kab-cja+cib}}-\\
&&\quad
T_{{2\,kab-cja-cib}}-T_{{2\,kab+cja+cib}},\\
f_0&=&2+T_{{2\,cja-2\,cib}}+T_{{2\,cja+2\,cib}}+T_{{4\,kab}}\\
&&\quad -T_{{2\,kab-2\,cib}}-T_{{2\,kab+2\,cib}}-T_{{2\,kab-2\,cja}}-T_{{2\,kab+2\,cja}}.
\end{array}
\end{equation}
If $c$ is even, we also have to compute the product of
$2 \varphi + 4 \cos\alpha \cos\beta$ in Formula (\ref{eq:rabc0}).
We write
$$
2 \varphi + 4 \cos\alpha \cos\beta
= \Frac 12 \Bigl (g_1(2\cos\frac{\pi}n) \varphi + g_0(2\cos\frac{\pi}n) \Bigr )
$$
where
\begin{equation}\label{eq:cf1}
g_1=2, \quad g_0=T_{{cja-cib}}+T_{{cja+cib}}.
\end{equation}
Using $T_{n+i} \equiv T_{n-i} \equiv -T_i \Mod{M_n}$, we can write
$f_0$, $f_1$ and $f_2$ as Chebyshev forms of degree at most $n-1$ and
$\taut(f_i) \leq 4$. We also have $\taut(g_i) \le 2$.

Using Corollary \ref{cor:prod2}, with
$k=N=\frac 12 (a-1)(b-1)(c-1)$ and $\tau=4$,
we see that we can compute the product $2^{2N} R_{a,b,c} \in
\ZZ[2\cos \frac{\pi}n][\varphi]$ in $\tcO(n^4)$ binary operations.
Each coefficient of $2^{2N} R_{a,b,c}$ has size
$\taut$ bounded by $\tcO(n)$, using Corollary \ref{cor:prod2}.
\EPf

%%%%%%%%%%%%%%%%%%%%%%%%%%%%%%%%%%%%%%%%%%%%%%%%%%%%%%%%%%%%%%%%%%%%%%%%%%%%%
\subsection{Computing $R_{a,b,c}$ by using numerical approximations}

One might use the expression of $R_{a,b,c}$ in
$\ZZ[2\cos \frac{\pi}n][\varphi]$ and approximate the coefficients of
$2^N R_{a,b,c}$ with accuracy less than $\frac{1}{2}$ in order to get
$R_{a,b,c}$ as an element of $\ZZ[\varphi]$:
using Corollary \ref{cor:brent2} below, it would take $\tcO(n^2)$
binary operations for each coefficient and
thus $\tcO(n^3)$ binary operations to get the entire polynomial.

We will improve this strategy by using numerical approximations
of the factors $4 \sin^2 \gamma \cdot P_{\alpha,\beta,\gamma}$ in $\QQ[\varphi]$.
We shall use the following technical lemma \citep[Lemma 18]{KRT15}
several times:
\begin{lem}\citep{Bren75,Bren76}\label{lem:brent1}
Let $0 \leq k \leq n$ and $\gamma  = 2 \cos k \frac{\pi}n$.
Let $\ell \in \ZZ_{>0}$.
One can compute $c \in \QQ$, of bitsize $\tau(c) \leq  \ell$ such that
$\abs{c - \gamma} \leq 2^{-\ell}$ in $\tcO(\ell + \log n)$ bit operations.
\end{lem}
From this Lemma, we deduce
\begin{cor}\citep[Cor.~19]{KRT15} \label{cor:brent2}% ancien approx1
Let $0 \leq k \leq n$ and
$\gamma  = 2 \cos k \frac{\pi}n$. Let $\ell \in \ZZ_{>0}$.
Let $f$ be the Chebyshev form $f_0 + \sum_{i=1}^{n-1} f_i T_i$ with
$\taut(f) \leq \tau$.
One computes $\widetilde F \in \QQ$ of bitsize $\tcO(n\tau+\ell)$
such that $\abs{\widetilde F - f(\gamma)} \leq 2^{-\ell}$ in $\tcO(n\ell + n\tau)$ bit operations.

One computes $F^-$ and $F^+$ of bitsize $\tcO(n\tau+\ell)$ such that
$F^- \leq F \leq F^+$ and
$F^+ - F^- \leq 2^{-\ell}$ in $\tcO(n\ell + n\tau)$ bit operations.
\end{cor}
We first show
\begin{lem}\label{lem:prodp1}
Let $P_i = a_i \varphi^2 + b_i \varphi + c_i$, $i=1,\ldots, N$,
be polynomials in $\RR_{\leq 2} [\varphi]$.
Let $\hat P_i \in \RR_{\leq 2}[\varphi]$, such that
$$
\norm{P_i}_{\infty} \leq M, \quad \norm{\hat P_i - P_i}_{\infty} \leq \delta .$$
Then we have
$$
\norm{\Prod_{i=1}^N P_i - \Prod_{i=1}^N \hat P_i }_{\infty} \leq
\delta\,  N \, 3^{N}(M+\delta)^{N}.
$$
\end{lem}
\Pf
It is straightforward that if $\deg Q \leq 2$ then
$\norm{PQ}_{\infty} \leq 3 \norm{P}_{\infty}\norm{Q}_{\infty}$.
We thus deduce by induction, with $\norm{\hat P_i}_{\infty} \leq M+\delta$, that
$\norm{P_1 \cdots P_k}_{\infty} \leq 3^{k-1} M^k$ and
$\norm{\hat P_1 \cdots \hat P_k}_{\infty} \leq 3^{k-1} (M+\delta)^k$.

Suppose that
$\norm{P_1 \cdots P_k - \hat P_1 \cdots \hat P_k}_{\infty} \leq \delta_k$, then
we obtain
\pn
$
\norm{P_1 \cdots P_{k+1} - \hat P_1 \cdots \hat P_{k+1}}_{\infty}$\\
\centerline{
$
\begin{array}{rcl}
&=& \norm{P_1 \cdots P_{k} (P_{k+1} - \hat P_{k+1}) +
(P_1 \cdots P_{k} - \hat P_1 \cdots \hat P_{k}) \hat P_{k+1}}_{\infty} \\
&\leq & \norm{P_1 \cdots P_{k} (P_{k+1} - \hat P_{k+1})}_{\infty}
+\norm{(P_1 \cdots P_{k} - \hat P_1 \cdots \hat P_{k}) \hat P_{k+1}}_{\infty} \\
&=& 3^{k} M^k \delta + 3\delta_{k} (M+\delta)
\end{array}$.}

We deduce that $\delta_{k+1} \leq 3 (M+\delta) \delta_k + (3M)^k \delta.$
Let $u_k = \Frac{\delta_k}{3^k (M+\delta)^k}$, we deduce that
$$
u_{k+1} - u_{k} \leq \Frac{\delta}{3(M+\delta)} \Bigl (\Frac{M}{M+\delta} \Bigr )^{k} \leq
\delta.
$$
We thus obtain that $u_{k+1} \leq u_{1} + k \delta= (k+1)\delta$.
\EPf
\begin{lem}\label{lem:prodp2}
Let $P_i = a_i \varphi^2 + b_i \varphi + c_i$, $i=1,\ldots, N$, be polynomials in
$\QQ_{\leq 2} [\varphi]$ with all coefficients being dyadic numbers
such that $\tau(P_i) \leq \tau$. Then
$P = \Prod_{i=1}^N P_i$ may be computed in $\tcO(N^2\tau)$ binary operations and
we have $\tau(P) \leq N\tau + (N-1)\log_2 3$.
\end{lem}
\Pf
From $\norm{U P_i}_{\infty} \leq 3 \norm{U}_{\infty} \norm{P_i}_{\infty}$, we obtain
that $\norm{P}_{\infty} \leq 3^{N-1} 2^{N\tau}$.
Let $m = \lceil \log_2 N \rceil$. We get $P_i = 1$ for $N<i\leq 2^m$ and
we compute by induction $P = \Prod_{i=1}^{2^m-1} P_i \cdot
\Prod_{i=1}^{2^m-1} P_{2^{m-1}+i}$, using fast multiplication in $\QQ[\varphi]$.

Let us suppose that we have computed $Q_0 = \Prod_{i=1}^{2^m-1} P_i$
and $Q_1 = \Prod_{i=1}^{2^m-1} P_{2^{m-1}+i}$
in $\tcO(2^{2m-2} \tau)$ binary operations.
$\tau(Q_1), \tau(Q_2) \leq 2^{m-1} (\tau + \log_2 3)$ and we obtain
$Q_1\cdot Q_2$  in $\tcO(2^{2m} \tau)$ binary operations.
At the end, we have computed $P$ in
$\sum_{i=1}^m \tcO(2^{2i} \tau) = \tcO(N^2\tau)$ binary operations.
\EPf
\pn
Applying the above results to the computation of $R_{a,b,c}$, we get:
\begin{prop}\label{prod:rabcnum}
Let $a$ and $b$ be coprime integers, $a$ odd, and $c$ an integer.
One can compute $R_{a,b,c}$ in $\tcO(n^3)$ binary operations.
\end{prop}
\Pf
Let $N= \frac 12 (a-1)(b-1)\pent c2$.
We have $\norm{4\sin^2\gamma P_{\alpha,\beta,\gamma}}_{\infty} \leq 16 - \frac{4}{c^2}=M$.
We compute each coefficient of $4\sin^2\gamma P_{\alpha,\beta,\gamma}$
with accuracy $\delta=2^{-6N+1}$. We obtain $N$ polynomials $\Pnulhat$ whose
coefficients are dyadic numbers of size bounded by $\tau = 6N+1$ and whose norm is bounded by $M+\delta \leq 16$.

We then compute $\Prod \Pnulhat$ in $\QQ[\varphi]$ in $\tcO(N^3)$ binary operations, using Lemma \ref{lem:prodp2} with $\tau = 6N+1$.

We have $\norm{R_{a,b,c} - \Prod \Pnulhat} \leq 2^{-6N+1}N 3^N \cdot 16^N
\leq \Frac 12$, using Lemma \ref{lem:prodp1}.
\EPf
%%%%%%%%%%%%%%%%%%%%%%%%%%%%%%%%%%%%%%%%%%%%%%%%%%%%%%%%%%%%%%%%%%

\section{Computing the real roots of $R_{a,b,c}$}\label{sec:compcrit}

In this section, the goal is to isolate efficiently the real
roots of $R_{a,b,c}$. As seen previously, the polynomial $R_{a,b,c}$
can be computed in
$\tcO(n^3)$ binary operations. Our first lemma gives estimates for the size of
the coefficients of $P_{\alpha,\beta,\gamma}$ such that
$\norm{R_{a,b,c}}_1 \leq 6^N$ and thus, running recent algorithms
such as in \citep{MS16}, the (real) roots of $R_{a,b,c}$ can be isolated in
$\tcO(n^3)$ binary operations.

In the next subsections, we will show that, in fact, the computation of
the real roots
of $R_{a,b,c}$ can be done in $\tcO(n^2)$ bit operations. This result
is due to many properties of the roots of $R_{a,b,c}$, in particular
the minimum distance
between two distinct real roots that is greater than $2^{-8n}$ (while
the worst case for a polynomial of degree $n$ with coefficients of
bitsize in $\tcO(n)$ is in $\tcO(n^2)$) and roots bounded in module by
$4$ (while the theoretical bound would have been in $\tcO(n)$).

Note that these two properties on the real roots could certainly be
used to adapt the complexity of the algorithm from \citet{MS16} or even
the one from \citet{RZ}, which has been used in \citep{KPR},  but we
will propose a dedicated algorithm for isolating the roots of $R_{a,b,c}$.

Let us start with the estimates for the size of the coefficients of
$P_{\alpha,\beta,\gamma}$ and $\norm{R_{a,b,c}}$:
\begin{lem}[Estimates]
Let $a$ and $b$ coprime integers, $a$ odd, and $c$ an integer. Let
$N=\frac 12 (a-1)(b-1)(c-1)$ and $R_{a,b,c} (\varphi)= \sum_{i=0}^N a_i \varphi^i$.
The we have $\norm{R_{a,b,c}}_1 = \sum_{i=1}^N \abs{a_i} \leq 6^N$.
\end{lem}
\Pf
According to \ref{rem:rabc12}, there are two cases to consider in Formula
(\ref{eq:rabc2}). If $c$ is odd then $R_{a,b,c}$ appears as the product
\[
R^{(1)}_{a,b,c} =
\Prod_{i=1}^{\frac{a-1}2}\Prod_{j=1}^{b-1}\Prod_{k=1}^{\frac{c-1}2}
4 \sin^2 \frac{k\pi}{c} P_{\frac{i\pi}{a},\frac{j\pi}{b},\frac{k\pi}{c}}(\varphi).
\]
If $c$ is even we have to multiply the previous product by
\[
R^{(0)}_{a,b,c} =
\Prod_{i=1}^{\frac{a-1}2}\Prod_{j=1}^{b-1} (2 \varphi + 4 \cos \alpha\cos\beta).
\]
Let $P = \sum_{i=0}^m a_i \varphi^i \in \RR[\varphi]$ and
$Q = \sum_{i=0}^m b_i \varphi^i \in \RR[\varphi]$.
We say that $P\prec Q$ if $\abs{a_i} \leq b_i$, $i=0,\ldots,m$.
It is straightforward that if $P \prec Q$ then $\norm{P}_1 \leq \norm{Q}_1$
and that if $P_1\prec Q_1$ and
$P_2 \prec Q_2$ then $\norm{P_1 P_2 }_1 \leq \norm{Q_1 Q_2}_1$.

We have
$\varphi+2\cos\frac{i\pi}{a}\cos\frac{j\pi}{b}\prec \varphi+2$, and
we deduce that $R^{(0)}_{a,b,c} \prec (2\varphi+4)^{(a-1)(b-1)/2}$.

If $\gamma = \frac{k\pi}2 \not = \frac{\pi}2$ then
\begin{eqnarray*}\label{eq:pabc}
\sin^2 \gamma P_{\alpha,\beta,\gamma}
&=&
\sin^2 \gamma \, \cdot \varphi^2+4 \varphi \cdot
\cos\alpha \cos\beta \sin^2 \gamma\\
&& \quad\quad\quad
+4 (\cos^2\alpha-\cos^2\gamma)(\cos^2\beta-\cos^2\gamma)\nonumber\\
&\prec& (\varphi+2)^2.
\end{eqnarray*}
We then deduce that $R^{(1)}_{a,b,c} \prec 
\Bigl( (2\varphi+4)^{2\pent{c-1}2} \Bigr)^{(a-1)(b-1)/2}$ and
$R_{a,b,c} \prec (2\varphi+4)^N$, that is
$\norm{R_{a,b,c}}_1 \leq 2^N \norm{(\varphi+2)^N}_1 = 6^N$.
\EPf
%%%%%%%%%%%%%%%%%%%%%%%%%%%%%%%%%%%%%%%%%%%%%%%%%%%%%%%%%%%%%%%%%%%%%%%
\subsection{Factor's roots}
Let $\alpha = \frac{i\pi}{a}$, $\beta = \frac{j\pi}{b}$ and
$\gamma =  \frac{k\pi}{c}$ with
$1 \leq i \leq \frac{a-1}2, \ 1 \leq j \leq b-1, \ 1 \leq k \leq \pent{c-1}2$.
\pn
If $\gamma=\frac\pi 2$, the unique root of $P_{\alpha,\beta,\frac{\pi}2}$ is
$-2\cos\alpha\cos\beta$.
If $\gamma \not = \frac{\pi}2$,
the discriminant of $P_{\alpha,\beta,\gamma}$ is
\[
\Delta_{\alpha,\beta,\gamma} =
16 \cos^2\gamma
\Bigl(1 -\Frac{\sin^2\alpha \sin^2 \beta }{\sin^2 \gamma}\Bigr).
\label{eq:ch5-Deltaijk}
\]
It has the same sign as
\begin{equation}
\sin \gamma -\sin\alpha \sin \beta,
\label{signdiscr}
\end{equation}
because $\sin \alpha, \sin\beta$ and $\sin \gamma$ are nonnegative.
The equation $\Delta_{\alpha,\beta,\gamma}=0$ is related to the equation
\begin{equation}
\sin r_1 \pi \sin r_2 \pi = \sin r_3 \pi \sin r_4 \pi
, \quad r_1, r_2, r_3, r_4 \in \QQ.
 \label{eq:sin4}
\end{equation}
All the solutions of Equation (\ref{eq:sin4}) are known:
\begin{lem}\citep{My,CJ}\label{lem:sin4}
Equation (\ref{eq:sin4}) admits
the one-parameter infinite family of solutions corresponding to
$$\sin \frac{\pi}6 \sin \theta =
\sin\frac{\theta}2 \sin (\frac{\pi}2 - \frac{\theta}2),$$
and a finite number of solutions listed in \citep{My}, for which the
denominators of the $r_i$ are not coprime.
\end{lem}
We thus deduce
\begin{prop}\label{prop:double1}
Let $\alpha = \frac{i\pi}a$, $\beta=\frac{j\pi}b$ and $\gamma=\frac{k\pi}c$,
where $(a,b)=1$ and $a$ is odd.
$P_{\alpha,\beta,\gamma}$ has a double root if and only if $\beta = \frac{\pi}2$ and
$\gamma = \alpha$. In this case, the double root is $\varphi=0$.
%\comm{PVK : c'est vrai $\varphi=0$!}
\end{prop}
\Pf
$P_{\alpha,\beta,\gamma}$ has a double root if and only if
$\Disc(P_{\alpha,\beta,\gamma})=0$,
that is to say $\sin\gamma\cdot 1 = \sin\alpha \sin\beta$. We conclude with the help of Lemma \ref{lem:sin4}.
\EPf
\pn
The knowledge of the sign of (\ref{signdiscr}) then gives explicit formulas
for the real roots of $P_{\alpha,\beta,\gamma}$ that we will explain in the
next section. But we also have to decide if the roots are distinct or what are their multiplicities.
%%%%%%%%%%%%%%%%%%%%%%%%%%%%%%%%%%%%%%%%%%%%%%%%%%%%%%%%%%%%%%%%%%%%%%%%%%%
\subsection{Multiple roots}
It may happen that $R_{a,b,c}$ has multiple real roots.
Two cases may occur: $P_{\alpha,\beta,\gamma}$
has a double root (that is $\Disc(P_{\alpha,\beta,\gamma})=0$) or
$P_{\alpha_1,\beta_1,\gamma_1}$ and $P_{\alpha_2,\beta_2,\gamma_2}$ have a common root
(that is $\Res_{\varphi} (P_{\alpha_1,\beta_1,\gamma_1},P_{\alpha_2,\beta_2,\gamma_2}) = 0$).

In the particular case when $\alpha_1=\alpha_2$ and $\beta_1=\beta_2$,
or $\gamma_1=\gamma_2=\frac{\pi}2$, the equation
$\Res_{\varphi} (P_{\alpha_1,\beta_1,\gamma_1},P_{\alpha_2,\beta_2,\gamma_2}) = 0$
may be solved using Lemma \ref{lem:sin4}.

\begin{prop}\label{prop:double2}
Let $\alpha_1 = \frac{i_1\pi}a$, $\beta_1=\frac{j_1\pi}b$,
and $\alpha_2 = \frac{i_2\pi}a$, $\beta_2=\frac{j_2\pi}b$,
where $(a,b)=1$. $P_{\alpha_1,\beta_1,\frac{\pi}2}$ and
$P_{\alpha_2,\beta_2,\frac{\pi}2}$ have a common root
if and only if $\alpha_1=\alpha_2$ and $\beta_1=\beta_2$.
\end{prop}
\Pf
The equation $\cos \alpha_1 \cos \beta_ 1 = \cos \alpha_2 \cos \beta_ 2$
admits the unique solution $\alpha_1=\alpha_2$ and $\beta_1=\beta_2$, using Lemma
\ref{lem:sin4}.
\EPf
\begin{prop}\label{prop:double3}
Let $\alpha = \frac{i\pi}a$, $\beta=\frac{j\pi}b$ and
$\gamma_1=\frac{k_1\pi}c$, $\gamma_2=\frac{k_2\pi}c$,
where $(a,b)=1$, $a$ is odd and $\gamma_1 \not = \gamma_2$. Then
$P_{\alpha,\beta,\gamma_1}$ and $P_{\alpha,\beta,\gamma_2}$ have a common
root $\varphi$ if and only if they are equal and one of the following cases occurs:
\bn
\item $\sin \alpha=\sin \gamma_1$, $\sin \beta = \sin \gamma_2$.
Their common roots are $\varphi=0$ and $\varphi = -4\cos\alpha\cos\beta$.
\item $\beta = \frac{\pi}6$, $\gamma_1 = \frac 12 \alpha$, $\gamma_2 =
\frac{\pi}2 -\alpha$.
In this case their common roots are $\varphi = -2\cos(\alpha \pm \frac{\pi}6)$.
\en
\end{prop}
\Pf
In this case $\Res(P_{\alpha,\beta,\gamma_1},P_{\alpha,\beta,\gamma_2})=0$, that is to say
\begin{equation}
(\sin^2 \gamma_1 - \sin^2 \gamma_2)
(\sin^2 \gamma_1 \sin^2 \gamma_2-\sin^2\alpha \sin^2 \beta)=0
\label{signresultant}
\end{equation}
We conclude using Lemma \ref{lem:sin4}.
\EPf
\pn
In case when $\alpha_1 \not = \alpha_2$ or $\beta_1 \not = \beta_2$,
%$P_{\alpha_1,\beta_1,\gamma_1}$ and $P_{\alpha_2,\beta_2,\gamma_2}$
%have a common root if
$\Res_{\varphi} (P_{\alpha_1,\beta_1,\gamma_1},P_{\alpha_2,\beta_2,\gamma_2})$
is equal to
$\Frac{\Delta_{1,2}}{D}$ where
$D=\sin^4 \gamma_1 \sin^4 \gamma_2$ and
\begin{equation}
\begin{array}{rcl}
\Delta_{1,2} &=& 16\left (
(\cos^2\alpha_1-\cos^2\gamma_1)
(\cos^2\beta_1-\cos^2\gamma_1)\sin^2\gamma_2 \right . - \\
&&\quad\quad\quad\left . \quad\quad (\cos^2\alpha_2-\cos^2\gamma_2)(\cos^2\beta_2-\cos^2\gamma_2)\sin^2\gamma_1\right )^2\\
&&\quad -4(\cos\alpha_1\cos\beta_1 - \cos \alpha_2 \cos \beta_2)\sin^2\gamma_1\sin^2\gamma_2 \times \\
&&\quad \left ((\cos^2\alpha_1-\cos^2\gamma_1)(\cos^2\beta_1-\cos^2\gamma_1)\cos \alpha_2 \cos \beta_2 \sin^2\gamma_2\right . -\\
&&\quad\quad\quad\left . (\cos^2\alpha_2-\cos^2\gamma_2)(\cos^2\beta_2-\cos^2\gamma_2)\cos\alpha_1\cos\beta_1 \sin^2\gamma_1\right ).
\end{array}\label{signresultant2}
\end{equation}
It would be interesting to get an arithmetic condition analogous to
Propositions \ref{prop:double1} and \ref{prop:double2},
asserting that $\Delta_{1,2}=0$.
We will see in the next section how we can
decide if $\Delta_{1,2}$ equals zero and if not, we can give an estimate of its
size.
%%%%%%%%%%%%%%%%%%%%%%%%%%%%%%%%%%%%%%%%%%%%%%%%%%%%%%%%%%%%%%%
\subsection{Bounds on roots}
The following result is an easy consequence of our results on Lissajous curves.
\begin{lem}
Let $\varphi$ be a root of $R_{a,b,c}$, then $\abs{\varphi}< 4$.
\end{lem}
\Pf
Let $\varphi$ be a root of $R_{a,b,c}$, then there exist $\alpha = \frac ia \pi$ and
$\beta = \frac jb \pi$ such that
$T_c (2 \cos (\alpha+\beta)+\varphi) =  T_c (2 \cos (\alpha-\beta)+\varphi)$. Using
Remark \ref{rem:inscribed}, we deduce that both $2 \cos (\alpha+\beta)+\varphi$ and
$2 \cos (\alpha-\beta)+\varphi$ belong to $[-2,2]$.
\EPf

We shall use the following lemma
\begin{lem}\label{lem:mincf}
Let $f = f_0 + \sum_{i=1}^{D} f_i T_i$
be an element of $\ZZ[t]$ expressed in
the Chebyshev basis. Then we have either $f(2\cos \frac{\pi}n)=0$ or
$\abs{f(2\cos \frac{\pi}n)} \geq \norm{f}^{(1-n/2)}_T$ where
$\norm{f}_T = \abs{f_0} + 2 \sum_{i=1}^{D} \abs{f_i}$.
\end{lem}
\Pf
Let $M_n$ be the minimal polynomial of $2\cos \frac{\pi}n$.
It is monic of degree $\frac 12 \phi(2n) \leq \frac n2$.
Using the $T_{n+i} \equiv T_{n-i} \equiv - T_{i} \Mod{M_n}$, we can write
$f \equiv \tilde f \;\mod M_n$ where $\deg \tilde f < n/2$ and
$\norm{\tilde f}_T \leq \norm{f}_T$.
%We can suppose now that $\deg f <n/2$.
\newcommand{\ndiv}{\hspace{-4pt}\not|\hspace{2pt}}%
We have $\Prod_{M_n(\gamma)=0} f(\gamma) = \Res(f,M_n)$.
If $f(2\cos \frac{\pi}n) \ne 0$ then $M_n \ndiv f$, and
$\abs{\Res(f,M_n)}\geq 1$ since $f$ and $M_n$ belong to $\ZZ[t]$.
When $\gamma_k = 2 \cos k \frac{\pi}n$ is a root of $M_n$ then
$\abs{f(\gamma_k)} \leq \norm{f}_T$ and we deduce that
$$
\abs{f(2\cos\frac{\pi}n)} \norm{f}^{\deg M_n - 1}_T \geq 1,
$$
which implies the announced result.
\EPf
%%%%%%%%%%%%%%%%%%%%%%%%%%%%%%%%%%%%%%%%%%%%%%%%%%%%%%%%%%%%%%%%%%%%%%%%%%
\begin{prop}[Discriminant]\label{prop:delta}
Let $\alpha=\frac{i\pi}a$, $\beta=\frac{j\pi}b$,
$\gamma=\frac{k\pi}c$ with $1\leq i\leq \frac{a-1}{2}$, $1\leq j\leq b-1$,
$1\leq k< \frac c2$. Then, either $\Delta_{\alpha,\beta,\gamma}=0$ or
$|\Delta_{\alpha,\beta,\gamma}|\geq 2^{-6n}$.

Moreover, if  $\beta \not =\frac{\pi}2$, then
$P_{\alpha,\beta,\gamma}$ has no double root and
there exists $\gamma_0 = k_0 \frac{\pi}c$ with $1 \leq k_0 \leq \pent c2$ such that
\begin{enumerate}
\item If $\gamma <\gamma_0$ then $\Delta_{\alpha,\beta,\gamma}<0$.
\item If $\gamma \geq \gamma_0$ then
$\Delta_{\alpha,\beta,\gamma}\geq 2^{-6n}$.%\Frac{1}{c^2} 24^{1-n/2}>0$
\end{enumerate}
\end{prop}
\Pf

From Proposition \ref{prop:double1}, $\Delta_{\alpha,\beta,\gamma}=0$
if and only if $\beta=\frac{\pi}2$ and $\alpha=\gamma$. Let $k_0 \leq \frac c2$ such that
$\sin(k_0-1)\frac{\pi}{c}< \sin\alpha\sin\beta \leq\sin\frac{k_0\pi}c$,
then $\Delta_{\alpha,\beta,\gamma} >0$ if and only if
$\gamma \geq \gamma_0 = \frac{k_0\pi}c$.

From Equation \ref{eq:ch5-Deltaijk}, If $\gamma \neq \frac{\pi}{2}$, then $\Delta_{\alpha,\beta,\gamma}=
\Frac{16}{\tan^2\gamma} \Bigl(\sin^2 \gamma -\sin^2\alpha \sin^2 \beta
\Bigr)$, otherwise, $P_{\alpha,\beta,\gamma}$ has a unique root.

In that case, as $\gamma \leq \frac{\pi}2 - \frac{\pi}{2c}$ then
$\tan^2 \gamma \leq 1/\cos^2 \gamma \leq 1/\sin^2 \frac{\pi}{2c}
\leq  c^2$, and thus
$
|\Delta_{\alpha,\beta,\gamma} |= \Frac{16}{\tan^2\gamma}
|\sin^2 \gamma - \sin^2 \alpha\sin^2 \beta |
\geq \Frac{1}{c^2} |f(2\cos \frac{\pi}n)|
$
where
$$
\begin{array}{rcl}
f(2\cos \frac{\pi}n) &=& 16 (\sin^2 \gamma - \sin^2 \alpha\sin^2 \beta)\\
&=&4-2\,\cos(2\alpha-2\beta) -2\,\cos(2\alpha+2\beta) +4\,\cos 2\beta + 4\,\cos 2\alpha -8\,\cos 2\gamma.
\end{array}
$$
Considering $f$ as the Chebyshev form
$$
f =
4-T_{{2\,jac-2\,ibc}}-T_{{2\,jac+2\,ibc}}+2\,T_{{2\,jac}}+2\,T_{{2\,ibc}}-4\,T_{{2\,kab}} ,$$
 we can see that $\norm{f}_T = 24$ so that, using Lemma \ref{lem:mincf},
 $|\Delta_{\alpha,\beta,\gamma}| \geq \Frac{1}{c^2} 24^{1-n/2}\geq 2^{-6n}.$
\EPf
%%%%%%%%%%%%%%%%%%%%%%%%%%%%%%%%%%%%%%%%%%%%%%%%%%%%%%%%%%%%%%%%%%%%%%%%%%
\begin{prop}[Separation] \label{prop:sep}
Let $a$, $b$ be coprime integers, $a$ odd, and $c$ be an integer.
Let $\varphi_1$ and $\varphi_2$ be two distinct real roots of $R_{a,b,c}$,
then $\abs{\varphi_1 - \varphi_2} \geq 2^{- 8n}$, where $n=abc$.
\end{prop}
\Pf
Consider two distinct roots $\varphi_1$ and $\varphi_2$
such that $P_{\alpha_1,\beta_1,\gamma_1}(\varphi_1)=0$ and
$P_{\alpha_2,\beta_2,\gamma_2}(\varphi_2)=0$.
Several cases may occur.
\begin{enumerate}
\item $P_{\alpha_1,\beta_1,\gamma_1}=P_{\alpha_2,\beta_2,\gamma_2}$.
It means that $2\cos\alpha_1\cos\beta_1=2\cos\alpha_2\cos\beta_2$
that is to say, because of the solutions of Equation (\ref{eq:sin4}),
$\alpha_1=\alpha_2=\alpha, \beta_1=\beta_2=\beta$. We thus have
$\gamma_1=\gamma_2$.
$\varphi_1$ and $\varphi_2$ are the two real roots of $P_{\alpha,\beta,\gamma}$ and we have
$$
\abs{\varphi_1-\varphi_2}^2 =
\Delta_{\alpha,\beta,\gamma} \geq 2^{-6n},%\Frac{\pi^2}{c^2} 24^{1-n/2},
$$
using Lemma \ref{prop:delta}.
\item
$\deg_{\varphi} P_{\alpha_1,\beta_1,\gamma_1}=\deg_{\varphi} P_{\alpha_2,\beta_2,\gamma_2}=1$, that is to say $\gamma_1=\gamma_2=\frac{\pi}2$.
We have
$$
\begin{array}{rcl}
\varphi_1 - \varphi_2 &=&
2\cos\alpha_1\cos\beta_1-2\cos\alpha_2\cos\beta_2\\
&=&\cos(\alpha_1+\beta_1)+\cos(\alpha_1-\beta_1)+\cos(\alpha_2+\beta_2)+\cos(\alpha_2-\beta_2)\\
&=& \frac{1}{2} f(2\cos\frac{\pi}{ab}).
\end{array}
$$
Here $f$ is the Chebyshev form
$T_{b\,i_1 + a\,j_1}+ T_{b\,i_1 - a\,j_1}+ T_{b\,i_2 + a\,j_2}+ T_{b\,i_2 - a\,j_2}$
with $\norm{f}_T \leq 4$.
We thus deduce that $\abs{\varphi_1-\varphi_2} \geq 4^{1-ab/2} = 2^{2-ab}$.
\item
$\deg_{\varphi} P_{\alpha_1,\beta_1,\gamma_1}=1, \deg_{\varphi} P_{\alpha_2,\beta_2,\gamma_2}=2$, that is to say
$\gamma_1=\frac{\pi}2 \not = \gamma_2$.\\
We have
$\varphi_1=-2\cos\alpha_1\cos\beta_1$,
$\varphi_2^{\pm}=-2\cos\alpha_2\cos\beta_2\pm \frac{1}{2}
\sqrt{\Delta_{\alpha_2,\beta_2,\gamma_2}}$.
Then
\begin{eqnarray*}
\abs{\varphi_1-\varphi_2^{\pm}} =
\abs{(\varphi_1-\varphi_2^{\pm})}
\Frac{\abs{\varphi_1-\varphi_2^{\mp}}}{\abs{\varphi_1-\varphi_2^{\mp}}}
&\geq&
\Frac{\abs{(\varphi_1-\varphi_2^{\pm})(\varphi_1-\varphi_2^{\mp})}}
{\abs{\varphi_1}+\abs{\varphi_2^{\mp}}}\\
&\geq& \Frac 18 \abs{(\varphi_1-\varphi_2^{\pm})(\varphi_1-\varphi_2^{\mp})}.
\end{eqnarray*}
But
$$
\begin{array}{rcl}
(\varphi_1-\varphi_2^{\pm})(\varphi_1-\varphi_2^{\mp}) &=&
(2\cos\alpha_1\cos\beta_1 - 2\cos\alpha_2\cos\beta_2)^2 - \Frac 14 \Delta_{\alpha_2,\beta_2,\gamma_2}\\
&=&\Frac 1{64 \sin^2 \gamma_2} f_2(2\cos\frac{\pi}n),
\end{array}
$$
where $f_2(2\cos\frac{\pi}n)= 64\sin^2\gamma_2 (2\cos\alpha_1\cos\beta_1 - 2\cos\alpha_2\cos\beta_2)^2 -
4\cos^2\gamma_2(\sin^2\gamma_2-\sin^2\alpha_2\sin^2\beta_2).$
We find that $\norm{f_2}_T = 256$ and then
$$
\abs{\varphi_1-\varphi_2^{\pm}} \geq \Frac{256}{8\cdot 64} 256^{-n/2} = 2^{-4n-1}.
$$
\item
$\deg_{\varphi} P_{\alpha_1,\beta_1,\gamma_1}=\deg_{\varphi} P_{\alpha_2,\beta_2,\gamma_2}=2$, that is to say $\gamma_1,\gamma_2\not =\frac{\pi}2$.
Let us suppose that $P_{\alpha_1,\beta_1,\gamma_1}$ and $P_{\alpha_2,\beta_2,\gamma_2}$
have a common root $\varphi$. \\
Then we have
$$
\varphi_1 - \varphi_2 = (\varphi + \varphi_1) - (\varphi+\varphi_2) = 2\cos\alpha_1\cos\beta_1-2\cos\alpha_2\cos\beta_2
$$
and we conclude that $\abs{\varphi_1 - \varphi_2} \geq \Frac{4}{2^{ab}}$.
\item
$\deg_{\varphi} P_{\alpha_1,\beta_1,\gamma_1}=\deg_{\varphi} P_{\alpha_2,\beta_2,\gamma_2}=2$, that is to say $\gamma_1,\gamma_2\not =\frac{\pi}2$.\\
Both $P_{\alpha_1,\beta_1,\gamma_1}$ and $P_{\alpha_2,\beta_2,\gamma_2}$ have two real roots $\varphi_1, \varphi'_1$ and $\varphi_2, \varphi'_2$.
These roots are distinct and their absolute values are bounded by 4.
We thus obtain
$$\Abs{\Res_{\varphi}(P_{\alpha_1,\beta_1,\gamma_1},P_{\alpha_1,\beta_1,\gamma_1})} =
\Abs{(\varphi_1 - \varphi_2)(\varphi_1 - \varphi'_2)(\varphi'_1 - \varphi_2)(\varphi'_1 - \varphi'_2)}
\leq 8^3 \abs{\varphi_1 - \varphi_2}
$$
We then obtain
$$
\abs{\varphi_1 - \varphi_2} \geq 2^{-9} \Abs{\Res_{\varphi}(P_{\alpha_1,\beta_1,\gamma_1},P_{\alpha_1,\beta_1,\gamma_1})}
= \Frac{\Delta_{1,2}}{\sin^4 \gamma_1 \sin^4 \gamma_2}.
$$
$256 \Delta_{1,2} = f_4(2\cos\frac{\pi}{n})$,
where $f_4$ is a Chebyshev form that satisfies $\norm{f_4}_1 \leq 32776 \leq2^{16}$.
We thus deduce that
$$
\abs{\varphi_1 - \varphi_2}\geq 2^{-17} 32776^{1-n/2} \geq 2^{-8n}.
$$
\end{enumerate}
We then obtain the announced result: $\abs{\varphi_1 - \varphi_2} \geq 2^{-8n}$.
\EPf
%%%%%%%%%%%%%%%%%%%%%%%%%%%%%%%%%%%
\subsection{Isolation}
We will rather compute independently the real roots of the polynomials
$P_{\alpha,\beta,\gamma}$ and compare them in order to get all the real
roots with their multiplicities. The first step is to compute the real roots of
$P_{\alpha,\beta,\gamma}$.
%%%%%%%%%%%%%%%%%%%%%%%%%%%%%%%%%%%%%%%%%%%%%%%%%%%%%%%%%%%%
\begin{lem}\label{lem:isolate1}
Let $a$ and $b$ be coprime integers and $c$ be an integer.
Let $\alpha=i\frac{\pi}a$, $\beta=j\frac{\pi}b$, $\gamma=k\frac{\pi}c$.
One can compute the real roots of $\Pnul$, if any, with precision $2^{-\ell}$
in $\tcO(\ell+n)$ binary operations.
\end{lem}

\Pf
The first operation consists in deciding if $\Pnul$ has $1$ double
real root $2$ simple real roots or $0$ real roots.

For the first case, it is just a matter of checking if
$\beta=\frac{\pi}{2}$ and $\gamma=\alpha$. In such a case, the unique
root is $-2\cos \alpha \, \cos \beta$ (Equation (\ref{eq:ch5-Deltaijk}))
which can be evaluated with precision
$2^{-\ell}$ in $\tcO(\ell)$ binary operations using \citep{Bren75,Bren76}.

Once we know that $\Pnul$ has no double root, checking the second and
third cases resume to computing the
sign of the discriminant $\Delta_{\alpha,\beta,\gamma}$ of
$P_{\alpha,\beta,\gamma}$, knowing that
$|\Delta_{\alpha,\beta,\gamma}|\geq 2^{-6n}$ (Proposition \ref{prop:delta}).

Deciding this sign can then be done by evaluating numerically
$\Delta_{\alpha,\beta,\gamma}$ with the precision $2^{-6n+1}$, which can
be performed in $\tcO(n)$ binary operations \citep{Bren75,Bren76}.

When $\Delta_{\alpha,\beta,\gamma}>0$, the roots can then
be computed as distinct roots of a quadratic univariate polynomial with
precision $2^{-6n+1}$ thanks to \citet{Bren75,Bren76}.
\EPf
\pn
We can now deduce directly the isolating intervals for the roots of $R_{a,b,c}$:
\begin{cor}\label{cor:roots1}
Let $a$ and $b$ be coprime integers and $c$ be an integer.
One can isolate the real roots of $R_{a,b,c}$ in intervals of length
$2^{-8n-1}$ in $\tcO(n^2)$ binary operations.
One can compute the real roots of $R_{a,b,c}$ and their multiplicities
in $\tcO(n^2)$ binary operations.
\end{cor}
\Pf
As already seen, $R_{a,b,c}$ is a product of $\cO(n)$ factors that are
either in the form $2\varphi+4 \cos\alpha \cos\beta$ or
$4 \sin^2 \gamma P_{\alpha,\beta,\gamma}$, when $\gamma \not = \frac{\pi}2$.

According to Proposition \ref{prop:sep}, the distance between two real
roots of $R_{a,b,c}$ is greater than $2^{-8n}$. Let us suppose all the
roots of all the factors have been computed independently with a precision
less than $2^{-8n+2}$, then two of these values approximate the same root if and
only if their difference is less than $2^{-8n}$.

Our first step thus consists in approximating all the roots of all the
factors of $R_{a,b,c}$ up to the precision $2^{-8n+2}$, which claims a
total of $\tcO(n^2)$ binary operations according to Lemma
\ref{lem:isolate1} for quadratic factors and \citep{Bren75,Bren76} for
linear ones.

The second step consists in sorting the list of approximations, which
claims $\tcO(n)$ comparisons between floating point numbers in
precision $\cO(n)$, say a total number in $\tcO(n^2)$ bit operations.

The final step consists in grouping the roots that are separated by a distance
less than $2^{-8n}$ which claims again $\tcO(n^2)$ binary operations.
\EPf

%%%%%%%%%%%%%%%%%%%%%%%%%%%%%%%%%%%%%%%%%%%%%%%%%%%%%%%%%%%%%%%%%%%%%%
\section{Computing the diagrams}\label{sec:diagrams}
We show here how we can decide if the curve $\cC(a,b,c,\varphi)$ is regular or
not. If it is regular, we show how we can determine its diagram, that
is to say the
signs of $Q_c(2 \cos (\alpha+\beta),2\cos(\alpha-\beta), \varphi)$,
for $\alpha=\frac{i\pi}a$ and $\beta=\frac{j\pi}b$.
We shall use the following lemma:
\begin{lem}\citep[Proposition 20]{KRT15} \label{lem:zero}
Let $f$ be a Chebyshev form of degree $d<n$ with $\taut(f)\leq \tau$.
Let $\gamma = 2 \cos k\frac{\pi}{n}$ where $(k,2n)=1$.
We can decide whether $f(\gamma)=0$ in $\tcO(n^2+n \tau)$ bit operations.
We can compute $\sign{f(\gamma)}$ in $\tcO(n^2 \tau)$ bit operations.
\end{lem}
%%%%%%%%%%%%%%%%%%%%%%%%%%%%%%%%%%%%%%%%%%%%%%%%%%%%%%%%%%
We then deduce:
\begin{lem}\label{lem:isolate2}
Let $a$ and $b$ be coprime integers and $c$ be an integer.
Let $\alpha=i\frac{\pi}a$, $\beta=j\frac{\pi}b$, $\gamma=k\frac{\pi}c$ and
$\varphi$ a rational number of bitsize $\tau$.
We can test if $\Pnul(\varphi)=0$ in $\tcO(n^2+n\tau)$ binary operations.
We can compute $\sign{\Pnul(\varphi)}$ in $\tcO(n^2\tau)$
binary operations.
\end{lem}
\Pf
Let $\varphi = \frac uv$.
When $\gamma=\frac{\pi}{2}$, we use Formula (\ref{eq:cf1}) and we get
$2 v P_{\alpha,\beta,\gamma} = 2u+4v \cos\frac{i\pi}{a}\cos\frac{j\pi}{b} =
f(2 \cos\frac{\pi}{ab})$, where $f =  vT_{{-ja+ib}}+vT_{{ja+ib}}+2\,u$.
Here $f$ is a Chebyshev form with
$\norm{f}_T \leq 4 \abs v + 2 \abs{u} \leq 6 \cdot 2^{\tau}$.
We thus obtain $\tau(u f) \leq 2\tau + \log_2 6= \cO(\tau)$.
The sign of $P_{\alpha,\beta,\gamma}(\varphi)$ is the sign of
$uf(2 \cos\frac{\pi}{ab})$.
\pn
If $\gamma = \frac{k\pi}2 \not = \frac{\pi}2$, then Formula (\ref{eq:cf2})
asserts that
$4 v^2 \sin^2 \gamma P_{\alpha,\beta,\gamma}(\varphi)=g(2 \cos\frac{\pi}n)$,
where
\begin{eqnarray*}
g &=& 2\,{u}^{2}+2\,{v}^{2}
-{u}^{2}T_{{2\,kab}}+{v}^{2}T_{{4\,kab}}
+2\,uvT_{{ibc-jac}}+2\,uvT_{{ibc+jac}}\\
&&\quad
-uv(T_{{ibc-jac-2\,kab}}+T_{{ibc+jac-2\,kab}}+
T_{{ibc-jac+2\,kab}}+T_{{ibc+jac+2\,kab}})\\
&&\quad
+{v}^{2}T_{{2\,ibc-2\,jac}}+{v}^{2}T_{{2\,ibc+2\,jac}}
-{v}^{2}T_{{2\,ibc-2\,kab}}-{v}^{2}T_{{2\,ibc+2\,kab}}\\
&&\quad\quad
-{v}^{2}T_{{2\,jac-2\,kab}}-{v}^{2}T_{{2\,jac+2\,kab}}
.
\end{eqnarray*}
Here $g$ is a Chebyshev form with
$\norm{g}_T \leq 4u^2+16 \abs{uv}+16v^2 \leq 36 \cdot 2^{2\tau}$.
We thus obtain $\tau(g) \leq 2 \tau + \log_2 36 = \cO(\tau)$.
The sign of $P_{\alpha,\beta,\gamma}(\varphi)$ is the sign of $g(2 \cos\frac{\pi}n)$.

We conclude using Lemma \ref{lem:zero}.
\EPf
%%%%%%%%%%%%%%%%%%%%%%%%%%%%%%%%%%%%%%%%
\begin{prop}
Let $a$ and $b$ be coprime integers and $c$ be an integer.
Let $\varphi$ be a rational number of bitsize $\tau$.
We can decide if $\cC(a,b,c,\varphi)$ is a knot in
running time $\tcO(n^2+n\tau)$, where $n=abc$.
We can compute the nature of the crossing points of $\cC(a,b,c,\varphi)$ in
$\tcO(n^2\tau)$ binary operations.
\end{prop}
\Pf
$\cC(a,b,c,\varphi)$ is a nonsingular curve if and only if $R_{a,b,c}(\varphi) \not =0$.
We first compute the $s$ real roots $\varphi_1, \ldots, \varphi_s$ of $R_{a,b,c}$
in isolating intervals of size $2^{-8n+1}$ in running time $\tcO(n^2)$,
using Corollary \ref{cor:roots1}.
Let us denote the isolating interval of $\varphi_i$ by $[u_{i},v_{i}]$, where
$u_{i} \leq \varphi_i \leq v_{i}$ and $\tau(u_i), \tau(v_i) \leq 8n+1$.
For each $i$ we assume that we know the list
of the $(\alpha,\beta,\gamma)$ such that $P_{\alpha,\beta,\gamma}(\varphi_i)$=0.
%This can be obtained in $\tcO(n^2)$ binary operations

We can find the unique $i_0$ such that $\varphi_{i_0} \leq \varphi < \varphi_{i_0+1}$ in
$\cO(s \log s)$ comparisons between $\varphi$ and the $u_{i}$'s. This claims
$\tcO(s (\tau + n)) = \tcO(n^2 + n\tau)$ binary operations.
\pn
Two cases may occur.
If $v_{i_0} < \varphi < u_{i_0+1}$ then $R_{a,b,c}(\varphi)\not = 0$ and
$\varphi_{i_0} < \varphi < \varphi_{i_0+1}$.

If $u_{i_0} \leq \varphi \leq v_{i_0}$, then we have to decide the sign of
$\varphi-\varphi_{i_0}$. Let us consider a polynomial
$P_{\alpha_0,\beta_0,\gamma_0}$ such that
$P_{\alpha_0,\beta_0,\gamma_0}(\varphi_{i_0})=0$.
$R_{a,b,c}(\varphi) = 0$ if and only if
$P_{\alpha_0,\beta_0,\gamma_0} (\varphi)=0$. This can be decided
in $\tcO(n^2+n\tau)$ binary operations, thanks to Lemma \ref{lem:isolate2}.

We can also compute the sign of
$P_{\alpha_0,\beta_0,\gamma_0}(\varphi)$ in $\tcO(n^2\tau)$ binary operations,
using Lemma \ref{lem:isolate2}.
If $\gamma_0 = \frac{\pi}2$ then clearly
$\varphi-\varphi_{i_0}$ and  $P_{\alpha_0,\beta_0,\gamma_0}(\varphi)$ have the same sign.

If $\gamma_0 \not = \frac{\pi}2$ then
$P_{\alpha_0,\beta_0,\gamma_0}(\varphi)$ has two roots: $\varphi_{i_0}$ and
$-2 \cos\alpha \cos \beta - \varphi_{i_0}$. Because
$\abs{\varphi - \varphi_{i_0}} < \cos\alpha \cos \beta$, we have
$(\varphi+2\cos\alpha\cos\beta)(\varphi-\varphi_0)P_{\alpha_0,\beta_0,\gamma_0}(\varphi) >0$.
We compute the sign of $\varphi+2\cos\alpha\cos\beta$ in $\tcO(n^2\tau)$ binary
operations and then deduce the sign of $\varphi-\varphi_{i_0}$ in $\tcO(n^2\tau)$
binary operations.
\pn
We have decided if $\cC(a,b,c,\varphi)$ is a knot in
$\tcO(n^2+n\tau)$ binary operations.
When $\cC(a,b,c,\varphi)$ is a knot, we have found $i_0$ such that
$\varphi_{i_0} < \varphi < \varphi_{i_0+1}$ in $\tcO(n^2 \tau)$ binary operations.
\pn
We have to determine the nature of the crossing over the $\frac 12 (a-1)(b-1)$
double points $A_{\alpha,\beta}$ of parameters
$(2\cos(\alpha+\beta), 2 \cos (\alpha-\beta))$ in
the plane curve $\cC(a,b)$.

Let  $\alpha=\frac{i\pi}a$ and $\beta=\frac{j\pi}b$. The real roots
$\varphi'_{1} = \varphi_{j_1} , \ldots, \varphi'_{k} = \varphi_{j_k}$  of
$Q_c(2\cos(\alpha+\beta), 2 \cos (\alpha-\beta))$
are selected within the roots of $R_{a,b,c}$ in $\tcO(n)$ binary operations.

We determine $k_0$ such that $\varphi'_{k_0} < \varphi < \varphi'_{k_{0}+1}$ in
$\tcO(\log k \log n) = \tcO(\log^2 n)$ binary operations, by inserting $i_0$
in the sequence $(j_1, \ldots , j_k)$.
The sign of $Q_c(2\cos(\alpha+\beta), 2 \cos (\alpha-\beta),\varphi)$ is then
$(-1)^{k_0}$. The sign of the crossing over the double point $A_{\alpha,\beta}$
is then $(-1)^{i+j+\pent{ib}a + \pent{ja}b + k_0}$ and may be computed
in $\cO(n)$ operations.
\pn
We have computed the nature of all the crossings in
$\cO(ab)\times \cO(n) = \cO(n^2)$ binary operations.
\EPf
\pn
We now show how we can list all the possible diagrams $\cC(a,b,c,\varphi)$.
\begin{prop}
Let $a,b,c$ be integers, $a$ is odd, $(a,b)=1$.
One can list all possible knots $\cC(a,b,c,\varphi)$ in
$\tcO(n^2)$ bit operations.
\end{prop}
\Pf
$\cC(a,b,c,\varphi)$ is a nonsingular curve if and only if
$R_{a,b,c}(\varphi) \not =0$.
We first compute the $s$ real roots $\varphi_1, \ldots, \varphi_s$ of $R_{a,b,c}$
in isolating intervals $[u_i,v_i]$ of size $2^{-8n+1}$ in running time $\tcO(n^2)$.

For each $i$ we assume that we know the list of the
$(\alpha,\beta,\gamma)$ such that $P_{\alpha,\beta,\gamma}(\varphi_i)$=0.

The knots $\cC(a,b,c,\varphi)$ are the same for $\varphi \in (v_k, u_{k+1})$
because the signs of the crossings over the double points $A_{\alpha,\beta}$
are constant.
For every $k$ in $1,\ldots, s$, we choose a rational number $r_{k}$ in
$(v_{k},u_{k+1})$. We choose $r_0 = -4$ and $r_{s+1}=4$.

For every  $\alpha$, $\beta$, we compute the
labels $(i_1 < \cdots < i_k)$ of the real
roots $\varphi'_{1} = \varphi_{i_1} , \ldots, \varphi'_{k} = \varphi_{i_k}$  of
$Q_c(2\cos(\alpha+\beta), 2 \cos (\alpha-\beta),\varphi)$
in $\tcO(n)$ binary operations.

Let $0 \leq k \leq s+1$.  We compute the sign of
$Q_c(2\cos(\alpha+\beta), 2 \cos (\alpha-\beta), r_k)$ in
$\cO(k)= \cO(c)$ binary operations.

The diagram of $\cC(a,b,c,r_k)$ is then determined in $\tcO(n)$
 binary operations.
\EPf

%%%%%%%%%%%%%%%%%%%%%%%%%%%%%%%%%%%%%%%%%%%%%%%%%%%%%%%%%%%%%%%
\section{Experiments}\label{experiments}

Our algorithms compute knot diagrams for Chebyshev space curves.
We do not discuss here the methods that are used to  identify the corresponding knot. They are based on the computation of polynomial knot invariants.

Following \citet*{KPR}, our first goal was to find the minimal 
Chebyshev parametrization for every two-bridge knot through 10 crossings,
that is to say $(a,b,c)$ minimal for the lexicographic order.

In  \citep{KPR}, resultants and Gr\"obner bases strategies were used for
computing the knot diagrams of $\cC(3,b,c,\varphi)$ and $\cC(4,b,c,\varphi)$ 
with $(b,c)<_{\mathrm{lex}}(14,300)$. Every two-bridge knot through 10
crossings was reached, except for six of them.

With the method we developed in the present article, we recover all the minimal
parametrizations from \citep{KPR} but also compute the six missing knots
parametrizations:
$$
\begin{array}{ll}
9_{5}=\cC(3,13, 326, 1/85),&
10_3 = \cC(4,13, 348, 1/138),\\
10_{30} = \cC(4,13, 306, 1/738),&
10_{33} = \cC(4,13,856,1/328),\\
10_{36}=\cC(3,14, 385, 1/146),&
10_{39}=\cC(3,14, 373, 1/182).
\end{array}
$$
From these results one deduces, for example, that there 
is no parametrization of $9_5$ as Chebyshev knot with $(a,b,c) <_{\mathrm{lex}} (3,13,326)$.
\pn
Some of the knots have parametrizations of high degree, which
explains that the straightforward strategies based on resultants
and/or Gr\"obner basis failed or took too much time. For example,
$R_{3,14,385}$ has degree 4992 and 2883 real roots which  are simple except 0 that is of multiplicity $6$.
$R_{4,13,856}$ has degree 15390 and 9246 real roots ($0$ has multiplicity 18).
We get 2050 non trivial knots, 83 of them are distinct,
and 63 have less than 10 crossings.
A table of these representations is posted on 
\href{https://team.inria.fr/ouragan/knots/}{https://team.inria.fr/ouragan/knots/}.
\pn
In these challenging experiments, a good strategy was to first try to
isolate separately the roots of the factors (of degrees at most $2$) of
$R_{a,b,c}$ using multiprecision interval arithmetic.
One has to notice that we did not use the theoretical separation bound
$2^{-8n}$, but a significantly lower precision was enough to separate the roots
of $R_{a,b,c}$.

The method we developed in this paper allows us to compute Chebyshev knot
diagrams for high values of $a$, $b$ and $c$.
Our experience with small $a$ and $b$ shows that the difficult cases
(multiple roots of $R_{a,b,c}$) we found were all predictable
(Prop. \ref{prop:double1},
\ref{prop:double2}, \ref{prop:double3}).
There are certainly some specific reasons connected with arithmetic
properties and the structure of cyclic extensions.
\pn
The main difference with the algorithm described in \citep{KPR} and the
computation of $R_{a,b,c}$ as a polynomial
of degree $\frac 12(a-1)(b-1)(c-1)$, is that it came as a resultant of
a polynomial of degree $(c-1)$ in $(X,\varphi)$ and a polynomial of
degree $\frac 12 (a-1)(b-1)$ in $X$ with coefficients in a unique
field extension.

Our computations can be considered as the extreme
case, in terms of degree, to be solved using methods
from the state of the art when running \citep{KPR} while it can be solved
in a few minutes with the method proposed in this article.
\pn
We consider that it might be one step further in the computing of polynomial
curves topology.

%\bibliographystyle{plain}

%\bibliographystyle{plain}
%\thispagestyle{empty}
%\bibliography{knot14}
\vfill
\hrule
\pn
P. -V. Koseleff, {\tt pierre-vincent.koseleff@imj-prg.fr}\\
UPMC-Sorbonne Universit\'es, Institut de Math\'ematiques de Jussieu (IMJ-PRG, CNRS 7586) and Ouragan INRIA Paris-Rocquencourt, France
\pn
D. Pecker, {\tt daniel.pecker@imj-prg.fr}\\
UPMC-Sorbonne Universit\'es, Institut de Math\'ematiques de Jussieu (IMJ-PRG, CNRS 7586), France
\pn
F. Rouillier, {\tt fabrice.rouillier@imj-prg.fr}\\
UPMC-Sorbonne Universit\'es, Institut de Math\'ematiques de Jussieu (IMJ-PRG, CNRS 7586) and Ouragan INRIA Paris-Rocquencourt, France
\pn
C. Tran, {\tt trancuong@hnue.edu.vn}\\
Department of Mathematics and Informatics, Hanoi National University of Education, Vietnam
\end{document}